\documentclass[aps,10pt,a4paper,showpacs,pre,twocolumn]{revtex4-1}
\usepackage{amssymb}
\usepackage{mathrsfs}
\usepackage{stmaryrd}
\usepackage{subfigure}
\usepackage{graphicx}
\usepackage{color}
\newcommand{\mb}{\mathbf}
\newcommand{\mc}{\mathcal}

\newcommand{\pdi}{\mbox{${N_w}/{N_n}$}}
\newcommand{\fig}{Fig.\ }
\newcommand{\eqn}{Eq.\ }
\newcommand{\eqns}{Eqs\ }
\newcommand{\RE}[1]{\textcolor{black}{#1}}

\begin{document}

\title{Polydisperse polymer brushes: internal structure, critical behavior, and interaction with flow}

\author{Shuanhu Qi}
\affiliation{Institut f\"{u}r Physik, Johannes Gutenberg-Universit\"{a}t Mainz, Staudingerweg 7, D-55099 Mainz, Germany}

\author{Leonid I. Klushin}
\affiliation{Department of Physics, American University of Beirut, P. O. Box 11-0236, Beirut 1107 2020, Lebanon}

\author{Alexander M. Skvortsov}
\affiliation{Chemical-Pharmaceutical Academy, Professora Popova 14, 197022 St. Petersburg, Russia}

\author{Friederike Schmid}
\affiliation{ Institut f\"{u}r Physik, Johannes Gutenberg-Universit\"{a}t Mainz, Staudingerweg 7, D-55099 Mainz, Germany}

\bigskip

\begin{abstract}

\textbf{ABSTRACT:} We study the effect of polydispersity on the structure of polymer brushes by
analytical theory, a numerical self-consistent field approach, and Monte Carlo
simulations. The polydispersity is represented by the Schulz-Zimm chain-length
distribution. We specifically focus on three different polydispersities
representing sharp, moderate and extremely wide chain length distributions and
derive explicit analytical expressions for the chain end distributions in these
brushes. The results are in very good agreement with numerical data obtained
with self-consistent field calculations and Monte Carlo simulations.  With
increasing polydispersity, the brush density profile changes from convex to
concave, and for given average chain length $N_n$ and grafting density
$\sigma$, the brush height $H$ is found to scale as \RE{
$(H/H_{\mathrm{mono}}-1)\propto(\pdi-1)^{1/2}$} over a wide range of polydispersity
indices $\pdi$ (here $H_{\mathrm{mono}}$ is the height of the corresponding
monodisperse brush.  Chain end fluctuations are found to be strongly
suppressed already at very small polydispersity.  Based on this observation, we
introduce the concept of the brush as a near-critical system with two
parameters (scaling variables), $(N_{n}\sigma^{2/3})^{-1}$ and
$(N_w/N_n-1)^{1/2}$, controlling the distance from the critical point.  This
approach provides a good description of the simulation data. Finally we study
the hydrodynamic penetration length $l_\mathrm{p}$ for brush-coated surfaces in
flow. We find that it generally increases with polydispersity.  The scaling
behavior crosses over from $l_\mathrm{p}\sim N_\mathrm{n}^{1/2}\sigma^{-1/6}$
for monodisperse and weakly polydisperse brushes to $l_\mathrm{p}\sim
N_{n}^{2/3}$ for strongly polydisperse brushes.

\end{abstract}

\maketitle

\newpage

\section{Introduction}

Polymer brushes have been the subject of intense investigations starting from
the seminal paper by Alexander roughly 40 years ago \cite{Alexander:1977}.
As surface modifiers, they have applications in various fields, e.g., 
as anti-fouling surfaces \cite{Biofouling}, lubrication
\cite{Klein:1991,Singh:2015}, biomedicine \cite{Zdyrko:2009}, \RE{biomaterial}
\cite{Ayres:2010}, nanosensoring \cite{Stimuli_responsive,Cohen_Stuart} and
others.  The current level of understanding of brushes is to a large extent 
based on a relatively simple but detailed analytical theory, which was
proposed at an early stage \cite{MWC,Zhulina,Skvortsov:1988}.  Although the
original theory described only equilibrium properties of monodisperse brushes,
it gave an impetus for a large body of experimental and simulation-based work
extending well outside this scope (see the reviews \cite{Klein:1991,
Advincula:2011,Binder_Kreer:2011,Halperin:1992}).  As far as the brush density
profiles and detailed characteristics of brush-forming chains are concerned,
the picture provided by theory and simulations for various chain lengths,
grafting densities and solvent qualities is extremely detailed. Thermodynamic
parameters such as the chemical potential and the osmotic pressure profiles are
also readily available \cite{Romeis:2012,Netz:2006,Netz:1998}.

The experimental synthesis of monodisperse brushes (or brushes with a very
narrow molar mass distribution) requires special methods. Polymer brushes
formed by long chains are typically produced using ``grafting to'' and
``grafting from'' techniques. In the ``grafting to'' technique, narrowly
fractioned pre-formed polymer chains of a given number-averaged chain length
$N_n$ are tethered to a surface either by chemical bonding or by strong
adsorption of a short sticky block. The
number of grafted chains per unit area, $\sigma$, can be calculated from the
total mass of grafted material and the total area, both directly measurable in
the experiment. During the brush synthesis, polymer molecules must diffuse 
through an existing grafted polymer layer to reach the reactive sites 
decorating the surface. Due to steric hindrance, this becomes increasingly
difficult with increasing height and density of the polymer brush. 
Hence only synthesize brushes with relatively low grafting densities 
and moderate brush thickness can be synthesized with this technique. \RE{Very recently,
it was reported in the literature \cite{Minko2016} that the highest grafting density
for poly(ethylene glycol) (PEG) brushes that would be obtained using the grafting to
method is close to 1.2 chains/nm$^2$.}

The ``grafting from'' approach has become the preferred option for the
synthesis of dense polymer brushes. This approach uses a surface immobilized
initiator layer and subsequent in situ polymerization to generate the polymer
brush. The method gives a polydisperse brush with higher grafting density
$\sigma$ independent of the polymerization time. In contrast to the ``grafting
to'' methodology, calculating $\sigma$ for systems prepared by the ``grafting
from" approach is more challenging because $N_{n}$ is unknown a priori. One
method of characterizing the chain length distribution (or the molecular weight
distribution) is to degraft the polymer chains from a large surface area and
collect a sufficient amount of material, which is then analyzed using a
sensitive analytical method of size exclusion chromatography. It was shown
\cite{Turgman} that the chain length distribution of the degrafted chains of
poly(methyl-methacrylate) is close to the so-called Schulz-Zimm distribution.
The values of the polydispersity index, $\pdi\thickapprox1.15-1.19$, were
marginally higher than those typically observed for bulk polymers grown under
similar conditions and did not reveal any significant dependence on the
polymerization time. (Here $N_w$ denotes the weight averaged chain length which
can be determined experimentally, e.g., by scattering methods). Thus the
experimental techniques available for producing polymer brushes with high
polymerization index typically yield rather broad chain length
distributions. Broad distributions are also typical for biological brushes
covering the surface of some cells or blood vessels \cite{Pandav}. 

It is common wisdom that polydispersity may significantly affect the properties
of polymeric systems. However, experimental work comparing the properties of
polymer brushes that differ only in polydispersity is still very rare.  In
Ref.~\cite{Balko:2013}, curves showing the energy vs.  separation upon bringing
together two brush-coated surfaces were compared  for brushes with
$\pdi\thickapprox1.06$ and $\pdi\thickapprox1.23$, and similar
values for the number-averaged molar mass $M_{n}=37000$ and the grafting
density $\sigma=0.195$.  Russell and co-workers \cite{Russel:2007} observed
that surfaces with a quaternary ammonia-modified dry polydisperse brush could
kill bacteria cells, while a polymer brush of the same thickness but with low
polydispersity could not penetrate the bacterial cell envelope (the charge
density is a critical parameter for the killing efficacy).

Polydispersity should also affect the penetration of polymer brushes by
proteins. In particular, differences in the fouling properties of brushes were
attributed to indirect polydispersity effects although no clear-cut
experimental evidence was given \cite{Krishnamoorthy:2014}.

From a theoretical point of view, the study of polydispersity effects on the
structure and properties of brushes is quite difficult.  A general approach
based on the self-consistent field (SCF) theory in the strong-stretching limit
was proposed by Milner, Witten, and Cates (MWC) \cite{Milner_Polydisp}. It
allows to calculate the brush density profile as well as individual chain
stretching characteristics numerically for a given chain length distribution.
Closed-form analytical expressions for the density profiles are typically not
available, although some global brush characteristics are given in a simple
integral form.  Intriguingly, almost none of the predictions have been compared
to numerical SCF work or simulations. At least part of the problem seems to be
psychological, since the variety of conceivable chain length distribution
functions is overwhelming. Whether the strong stretching approximation which
proved to be so useful in the case of monodisperse brushes is practical for
broad chain length distributions remains an open question.  Monte Carlo (MC)
simulations of polydisperse brushes face the additional challenge that one must
perform quenched disorder averages, due to the fact that chains of different
length are permanently grafted to a substrate. An extensive numerical SCF study
of polydispersity effects on the brush structure was presented in
\cite{Leermakers} on the basis of the Schulz-Zimm distribution, which is
commonly used to describe experimental samples. With this choice of
distribution shape, the polydispersity parameter $\pdi$ can be tuned in a range
from 1 to 2. Brush density profiles for moderate grafting density and good
solvent conditions were calculated, allowing to study systematic changes in the
brush height and in the profile shape with increasing polydispersity.
Qualitatively, the picture can be summarized as follows: Starting from the
familiar parabolic (convex cap-like) profile in the purely monodisperse case,
the mean curvature of the profile decreases monotonically, resulting in clearly
concave shapes at $\pdi>1.4$. In the case of moderate polydispersity,
$\pdi\simeq1.15$, the density profile is almost linear. Another paper by the
same group addressed the polydispersity effects on the brush penetration by
nanoparticles using SCF calculations \cite{deVos_nanopartcle}.  It was shown
that the larger the polydispersity, the easier it is for a small particle to
penetrate the brush and to touch the substrate. For very large particles an
opposite effect is found: it is harder to compress a polydisperse brush than a
corresponding monodisperse brush.

In the present paper we study three cases representing vanishingly small,
moderate and large polydispersity. By combining closed-form analytical
solutions based on the MWC theory with a Green's function approach we
calculate the density profiles and the individual chain conformations
(including the chain fluctuations neglected by the strong stretching
approximation) and test the theory predictions by MC simulations and
one-dimensional SCF numerics.  Finally, we examine the effect of
polydisperse brushes on hydrodynamical shear flow and calculate in particular
the hydrodynamic penetration depth.

The remainder of the paper is organized as follows: Sec.\ \ref{sec:model}
describes the model system, the MC scheme, and the Schulz-Zimm distribution. In
Sec.\ \ref{sec:height}, we discuss general characteristics of polydisperse
brushes, such as their height and their density profiles.  In Sec.\
\ref{sec:theory_tools}, we briefly summarize the main predictions of the
analytical theory (which are derived in detail in Appendix
\ref{sec:theory-appendix}, based on the MWC theory).  The theoretical
predictions are compared with MC simulations and numerical SCF calculations in
Sec.\ \ref{sec:results}. In this context, we also analyze the chain end
fluctuations in weakly polydisperse brushes and introduce the concept of
monodisperse brushes as near-critical systems.  The implications for the
interaction with hydrodynamic flows are discussed in Sec.\
\ref{sec:hydrodynamics}.  We close with a summary of the main features of
polydisperse brushes in Sec.\ \ref{sec:summary}.  The Appendix summarizes
technical details on the SCF theory and presents the derivations of analytical
results -- in particular, the analytical expressions for the Green's function
of single chains embedded in three types of brush (monodisperse, moderately
polydisperse with exactly linear density profile, and strongly polydisperse)
and the expressions for the hydrodynamic penetration length in these brushes.

\section{Model and Monte Carlo scheme}\label{sec:model}

The system is composed of a dense polydisperse brush in implicit solvent in a
volume $V=L_x\cdot L_y\cdot L_z$. We adopt periodic boundaries along the $x$
and $y$ directions, while impenetrable boundary walls are placed $z=0$ and
$z=L_z$. Polymer chains are modelled as chains of beads connected by harmonic
springs with the spring constant $3k_\mathrm{B}T/2a^2$, where $a$ is the
statistical bond length, $k_B$ is the Boltzmann constant, and $T$ the
temperature. We will use $a$ as the unit length and $k_BT$ as the energy unit.
Brush chains in the system are grafted with one end each onto a flat substrate
located at $z_0$, which is chosen $0<z_0<a$ for practical reasons. The
nonbonded interaction is formulated in terms of the local monomer density, 
implying that we have soft interactions. The Hamiltonian thus
has the Edwards type \cite{EdwardsType1, EdwardsType2, EdwardsType3,
Hybrid_PF, Klushin:2015} and can be written as
\begin{eqnarray}
\beta\mc H&=&\frac{3}{2a^2}\sum_{\alpha=1}^{n_b}\sum_{j=1}^{N_j-1}\Big(\mb R_{\alpha j}-\mb R_{\alpha, j-1}\Big)^2\nonumber\\
&&+\frac{v}{2}\int d\mb r\hat\phi^2(\mb r)
\end{eqnarray}
where $\beta\equiv1/k_BT$, and the bond length $a$ is assumed to be the same
for each monomer. In the first term $\mb R_{\alpha j}$ denotes the location of
the $j$th bead in the $\alpha$-th chain, the index $\alpha$ running over all
$n_b$ brush chains, and $N_\alpha$ is the chain length for the $\alpha$-th
brush chain.  The second term represents the nonbonded effective
monomer-monomer interactions. Here, $v$ is the excluded volume interaction
parameter measuring the monomer-monomer interaction strength in good solvent.
The microscopic density of monomers is a function of conformation,
$\hat\phi=\sum_{\alpha=1}^{n_b}\sum_j\delta(\mb r-\mb R_{\alpha j})$. We will
use $\phi$ to denote the smoothed density, for example, the ensemble average
density, i.e., $\phi\equiv\langle\hat\phi\rangle$.  In the present work, we
choose $v=1$, which corresponds in explicit solvent to a Flory-Huggins
parameter $\chi=0$ (athermal solvent conditions) \RE{in mean-field approximation \cite{chi_v}.}  Furthermore, we set
$\beta=1$ and $a=1$, thus defining the simulation units for length and energy.

In the MC simulations, local densities are extracted from the position of the
beads using a Particle-to-Mesh technique \cite{Particle_Mesh}, which provides a
way for the smoothing of density operators. In the present work we use the
first order Cloud-in-Cells scheme \cite{CIC}, which means that monomers are
partitioned between their eight nearest neighbour vortices with a weight that
depends on the relative distance between the monomer and the vortex. This
implies that two beads interact even when they are in neighbour cells, and thus
the interaction distance is effectively larger than cell size. Higher order
schemes are conceivable, but computationally more expensive.

For the chain length distribution, we choose the Schulz-Zimm (SZ) distribution
\cite{Schulz,Zimm}, a realistic size distribution that is often adopted to
describe polymer polydispersity. The continuous SZ distribution is a
two-parameter function, and if one uses $N_n$, the number-averaged chain length
and $k$, a parameter related to the polydispersity index, as the free
parameters, the SZ distribution can be written as
\begin{equation}
P(N)=\frac{k^kN^{k-1}}{\Gamma(k)N_n^k}\exp\Big(-k\frac{N}{N_n}\Big),
\end{equation}
where $\Gamma(k)$ is the Gamma function. In the limit of $k\to\infty$ the
distribution becomes a $\delta$-peak, while for $k=1$, it reduces to a simple
exponential distribution. It is easy to check that $P(N)$ is normalized to
unity, and the first order moment is given by $\langle N\rangle=\int_0^\infty
dNNP(N)=N_n$ as expected, while the high order moments can be obtained by a
recursion formula $\langle N^m\rangle=\langle N^{m-1}\rangle
N_n\frac{k+m-1}{k}$. The weight-averaged chain length can be calculated as
$N_w\equiv\langle N^2\rangle/\langle N\rangle=N_n\frac{k+1}{k}$. The relation
between $N_w$ and $N_n$ can be expressed in terms of $k$ as $\pdi=1+1/k$.

In practice we have only a finite number $n=\sigma\cdot L_x \cdot L_y$ of
chains in the system ($\sigma$ is the grafting density).  We determine all
possible sets of $n$ chains whose first, second, and third chain length moment
are consistent with those of the SZ distribution within one percent.  Then we
perform SCF calculations to estimate the brush density profiles of each set
(they are almost identical) and select the set that produces the smoothest
profiles. This set is then used for the simulations. Disorder averages are
carried out with respect to the arrangement of grafting points of the chains \cite{SCF_book}.

In the present study, the number averaged chain length is fixed at $N_n = 100$.
The system is divided into cubic cells with the size $a^3=1$.  To study the
effect of polydispersity, we specifically choose three different
polydispersities with $k=50, 7$, and 1 representing vanishingly small, moderate
and large polydispersity, and study brushes with grafting density $\sigma=0.1,
0.2, 0.3$. \RE{At chain length $N=100$, the mean squared radius of gyration of free chains is $R_g^2=N/6$.
The overlap area density is calculated as $\sigma^*=\frac{1}{\pi R_g^2}\simeq 0.02$.
Therefore the crowding $\sigma/\sigma^*$ at our grafting densities ranges from 5 to 15.}
The grafting points of the chains on the substrate are fixed on a
regular square lattice. The system size is chosen $L_{x}=60$, $L_{y}=60$, and
$L_{z}=200$.

These systems were studied by MC simulations. In every MC update, we try to
move the position of one chosen monomer to a new position with a distance in
space comparable to the bond length.  This trial move results in an energy
change including the bonded and non-bonded energy, and it is accepted or
rejected according to the Metropolis probability. In all simulations,
$4\times10^{5}$ MC steps per monomer were performed to equilibrate the system,
and another $4\times10^{5}$ MC steps to extract statistical averages. 
\RE{We checked the equilibration by running further simulations up to $8\times 10^5$
MC steps. Consistent results were obtained for both brush properties and single chain behavior.
}
Final statistical quantities were obtained by
averaging the results from 48 separate independent MC runs with different
arrangement of the grafting points.

\section{General characteristics of polydisperse brush: height and density}\label{sec:height}

\begin{figure}[t]
\centerline{\includegraphics[width=7cm]{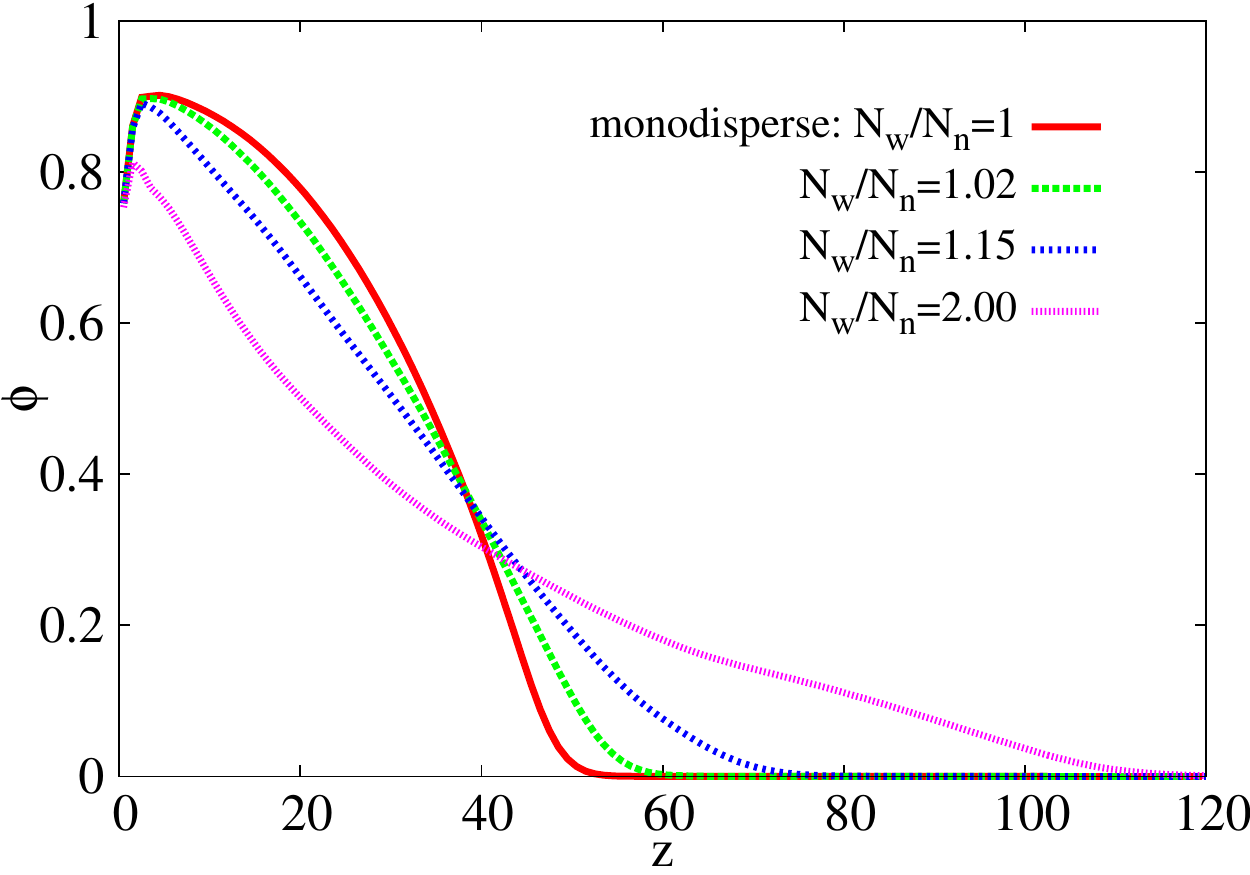}}
\caption{Density profile $\phi(z)$ of polymer brushes at fixed number average
$N_{n}=100$ and $\sigma=0.3$ for several polydispersity indices 
$\pdi=1,1.02,1.15,2$ as indicated. }
\label{fig:profiles} 
\end{figure}

A family of brush density profiles for brushes with Schulz-Zimm chain length
distributions and polydispersity indices $\pdi=1,\:1.02,\:1.15,\:2$ at fixed
surface grafting density $\sigma=0.3$ and number average $N_{n}=100$ is shown
in \fig \ref{fig:profiles}.

As the polydispersity is increased, the shape of the brush density profiles
changes from a concave parabola to a convex curve, and the brush height
increases. Several observations turn out to be of very general validity. First,
the area under all the profile curves is the same. This follows from the 
definition of the number averaged chain length, $N_{n}$. Indeed, the area under
the curve corresponds to the total number of monomers (per unit grafting
area), which is the product of the number of chains per unit area, $\sigma$,
and the average number of monomers per chain, $N_{n}$.  Thus
$\int\phi(z)dz=\sigma N_{n}$ irrespective of the particular shape of the chain
length distribution. The next observation concerns the local brush density near
the solid substrate. At the surface one observes a small dip which is of the
order of a few monomer sizes and may be model-dependent.  It is known that
off-lattice molecular chain models may produce density oscillations very close
to the surface as a result of packing effects \cite{Binder_Milchev_2012}. Here,
we are interested in the extrapolated density at the surface, $\phi(0)$, as
the property of the smooth model-independent profile. It is clear for \fig
\ref{fig:profiles} that $\phi(0)$ changes little with increasing
polydispersity index, although some decrease is observed for the largest
polydispersity, $\pdi=2$. The analytical theory of polydisperse brushes
developed by Milner, Witten, and Cates (MWC) predicts that in the
strong-stretching limit under good solvent conditions, within the second-order
virial expansion approximation, the extrapolated density $\phi(0)$ is
completely determined by one single parameter: the product of $\sigma$ and the
second virial coefficient $v$. The same holds for chain length
distributions of any shape and any average chain length, and hence, the
expression derived for monodisperse brushes is very generally valid:
\begin{equation}\label{eq:fi0}
\phi(0)=\phi_{0}=\frac{3}{2}\left(\frac{\pi\sigma v}{2}\right)^{2/3}
\end{equation}

Most experimental studies of polymer brushes have focused on investigating
the brush height as a function of the average chain length and the
grafting density. The original Alexander-de Gennes theory predicts
the scaling form $H\sim N\sigma^{1/3}$and the more detailed SCF theory
of monodisperse brushes \cite{MWC, Zhulina} gives the same scaling
with a specific prefactor: 
\begin{equation}\label{eq:Hmono}
H_{\mathrm{mono}}=\left(\frac{4\sigma v}{\pi^{2}}\right)^{1/3}N_{n}
\end{equation}

Numerical SCF calculations \cite{Leermakers} have indicated that the same
scaling applies to polydisperse brushes.  In Appendix \ref{sec:theory-appendix}
we demonstrate that this scaling follows directly from the MWC theory for any
realistic chain length distribution. Thus the form of the chain length
distribution only affects the prefactor. For any single-parameter family of
chain length distribution the pre-factor becomes a function of the
polydispersity index, hence the height can be written as $H=A\left(\pdi \right)
\: N_{n}\sigma^{1/3}.$ Our goal is to estimate the pre-factor $A(\pdi)$ from
our MC simulations. Unfortunately, extracting the brush height from numerical
data on the density profiles is not entirely straightforward because of the
very tenuous tail caused by the chain fluctuations. In monodisperse
brushes, a popular way to overcome this problem is to simply extrapolate the
parabolic profiles, but this approach is not applicable in general polydisperse
brushes.  Hence we adopt a simple pragmatic approach and define the brush
height $H$ as the distance where the density becomes smaller than a threshold
value of 0.01.

\begin{figure}[t]
\centerline{\includegraphics[width=7cm]{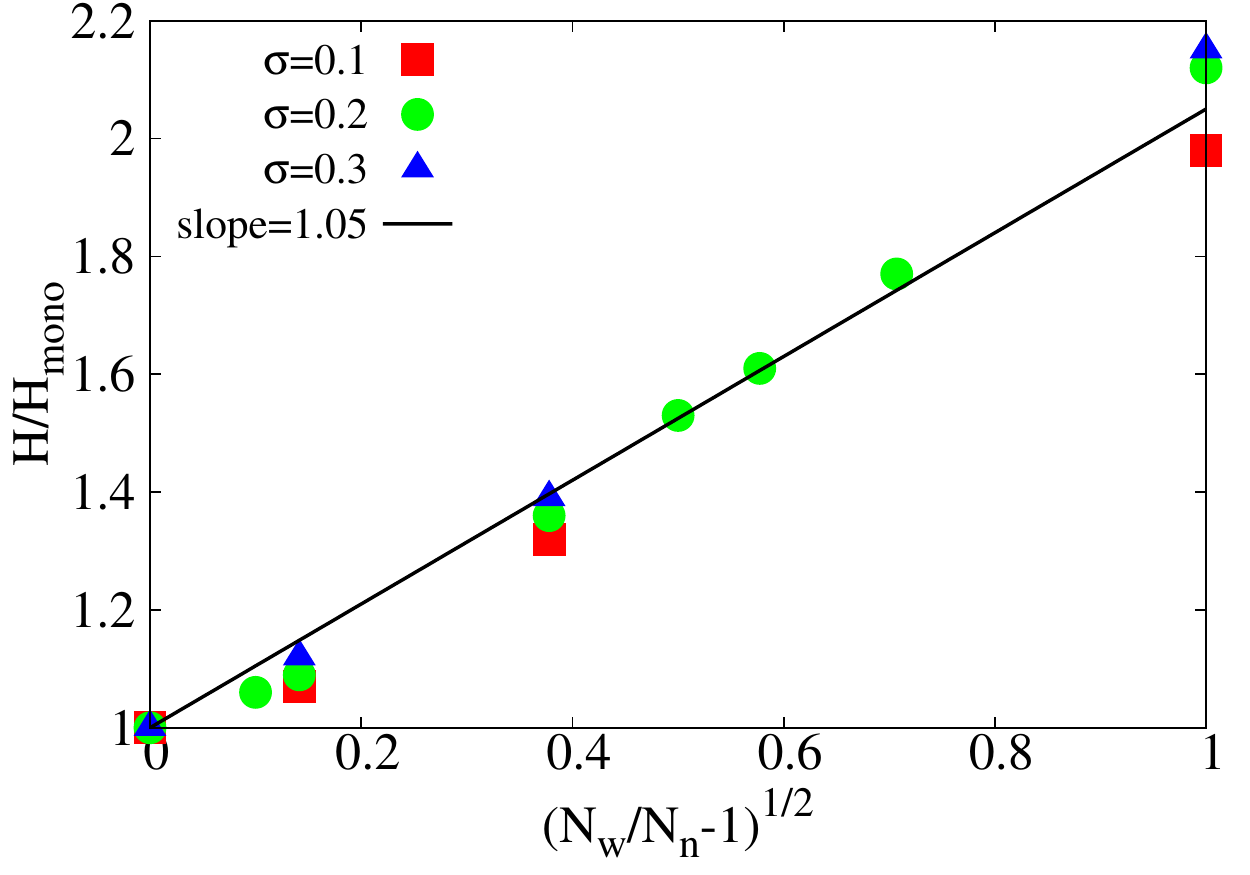}}
\caption{Brush height of polydisperse brushes $H(N_{n},\sigma,k)$ reduced
by the height of the corresponding monodisperse brush 
$H_{\mathrm{mono}}(N_{n},\sigma)$ as a function the polydispersity parameter 
$(\pdi-1)^{1/2}$ at three different grafting densities $\sigma=0.1,$ 
0.2 and 0.3. The straight line corresponds
to \eqn (\ref{eq:relative_brush_thickness}) }
\label{fig:brush_heights} 
\end{figure}

Analytical considerations for specific chain length distributions
\cite{Milner_Polydisp,Klushin:1992} have suggested that the average
height of a polydisperse brush $H$, rescaled by the height of the
corresponding monodisperse brush $H_{\mathrm{mono}}(N_{n},\sigma)$,
should be a linear function of $(\pdi-1)^{1/2}$ for narrow
chain length distributions. We test our simulation results for the brush height against this theoretical
prediction. For the SZ distribution, we have $(\pdi-1)=1/k$, and hence
we plot the ratio $H(N_{n},\sigma,k)/H_{\mathrm{mono}}(N_{n},\sigma)$
in \fig \ref{fig:brush_heights} vs. $(\pdi-1)^{-1/2}$ for different grafting 
densities $\sigma$. \RE{The data roughly collapse onto a single straight line,
which can be described by a simple equation
\begin{equation}\label{eq:relative_brush_thickness}
H/H_{\mathrm{mono}}=1+\alpha\left(\pdi-1\right)^{1/2}
\end{equation}over a wide range of polydispersity indices. Here $\alpha\approx 1$ is an empirical prefactor
whose exact value depends on the cut-off value used for estimating the brush height. For example, if the cut-off value
is chosen 0.01, we get $\alpha=1.05\pm0.03$ as shown in Fig.(\ref{fig:brush_heights});
for a cut-off value 0.001, we get $\alpha=1.15\pm0.03$ (data not shown).
At low polydispersity, the brush height is almost independent of the predefined cut-off,
since the brush tail is sharp; at high polydispersity, when the brush tail becomes flat,
the estimated brush height is sensitive to the cut-off.}

\section{Analytical results}\label{sec:theory_tools}

Explicit analytical predictions for the structure of polymer brushes with
experimentally relevant chain length distributions are not found in the
literature. Here we provide some results for three cases inspired by simulation
results for brushes with SZ polydispersity. The brush density profiles
presented above in \fig \ref{fig:profiles} (which correspond to moderate
grafting densities and good solvent conditions) illustrate the systematic
changes of the profile shape with increasing polydispersity. For weakly
polydisperse brushes with $\pdi \simeq1.02$, the profiles change very little
compared to the purely monodisperse parabolic brush. Thus we can apply exact
results available for the latter case to identify weak polydispersity effects.
The moderately polydisperse case is exemplified by the density profile which
has a nearly constant slope for $\pdi \simeq1.15$. According to the basic
tenets of the SCF theory, a linear density profile generates a linear
mean-field potential. Fortunately, exact solutions are also available for
chains in a linear external potential \cite{Mansfield}.  Although for the SZ
case the profile is only approximately linear, we demonstrate that a somewhat
different chain length distribution with a similar polydispersity index
generates a strictly linear profile (at least in the strong stretching limit).
Finally, the case of the largest polydispersity with an exponential chain
length distribution and $\pdi \simeq2$ also admits an exact analytical solution
in the framework of the MWC theory.

We analyze not only the average brush properties such as the density profile,
but also the behavior of individual brush chains including the mean end
positions and their fluctuations. In the following, we list the most important
fundamental analytical results.  All derivations and most expressions that are
used later for comparisons with the simulations data are found in Appendix
\ref{sec:theory-appendix}.

\subsection{Monodisperse polymer brush}

The density profile of monodisperse brushes is given by 
a well-known expression: 
\begin{equation}\label{eq:mono_profile-1}
\phi(z)=\phi_{0}\left(1-\frac{z^{2}}{H_{\mathrm{mono}}^{2}}\right)
\end{equation}with $\phi_{0}$ and $H_{\mathrm{mono}}$ given by \eqns 
(\ref{eq:fi0}) and (\ref{eq:Hmono}). The mean-squared 
fluctuations of the chain end position are also known:
\begin{equation}\label{eq:mono_end_fluct}
\mathrm{var}Z_{\mathrm{mono}}=\left(\frac{2}{5}-\left(\frac{3\pi}{16}\right)^{2}\right)H_{\mathrm{mono}}^{2}
\end{equation}

In the polydisperse brushes, our main interest will be to investigate the
end-monomer distributions for chains of different length. To enable a
meaningful comparison, we introduce a vanishing fraction of minority chains of
different length, $N$, into the otherwise monodisperse brush. The end monomer
probability density for a minority chain is deduced from the Green's function
which is a known solution of the Edwards' equation in the presence of a purely
parabolic potential and an impenetrable inert grafting surface
\cite{Skvortsov:1997}: 
\begin{equation}\label{eq:Green_parabolic-1}
G(z,N)=B\: z\exp\left[-\frac{3\pi}{4N_{n}}\cot\left(\frac{\pi N}{2N_{n}}\right)z^{2}\right]
\end{equation}where $B$ is a normalization factor. 

\subsection{Moderately polydisperse brush}

The moderately polydisperse brush considered here is characterized by a
linear density profile: 
\begin{equation}
\phi_{\mathrm{lin}}(z)=\phi_{0}-fz\label{eq:linear_profile-1}
\end{equation}
where $\phi_{0}$ is given by \eqn (\ref{eq:fi0}). The brush height is: 
\begin{equation}
H=\frac{4}{3}H_{\mathrm{mono}}.
\end{equation}and the mean force per monomer is $f=\phi_{0}/H$. The Green's 
function of a chain placed in a uniform force field of strength $f$ is a 
solution of the corresponding Edwards equation which was obtained in 
Ref.\ \cite{Mansfield}:
\begin{equation}\label{eq:Green_function_at_k7}
G(z,N)=\left(\frac{2\pi N}{3}\right)^{-1/2}\exp\left[\frac{f^{2}N^{3}}{18}-\frac{3}{2N}\left(z-\frac{f\, N^{2}}{6}\right)^{2}\right]
\end{equation}

On the basis of the MWC theory, one can 
find a closed-form analytical solution for the chain length distribution
$P_{\mathrm{lin}}(N)$ that generates a brush with exactly linear
profile (see Appendix \ref{sec:theory-appendix}):
\begin{equation}\label{eq:CLD_linear_profile-1}
P_{\mathrm{lin}}(N)=\frac{3N}{N_{\mathrm{max}}^{2}}\left[1-\frac{N^{2}}{N_{\mathrm{max}}^{2}}\right]^{1/2}
\end{equation}
where the cut-off chain length is $N_{\mathrm{max}}=\frac{16}{3\pi}N_{n}$.
The polydispersity index of this distribution is $\pdi =1.15$, which is
close to the value $\pdi=1.14$ of the SZ distribution producing
approximately linear brush density profile.

\subsection{Strongly polydisperse brush}
For the strongly polydisperse case with $\pdi=2$, the SZ distribution 
simplifies to a purely exponential form: 
\begin{equation}\label{eq:ZimmShulz_at_k1-1}
P(N)=\frac{1}{N_{n}}\exp\left(-\frac{N}{N_{n}}\right)\ 
\end{equation}
which allows an exact closed-form solution on the basis of the MWC theory.

The brush density profile is given in the form 
\begin{equation}
\phi_\mathrm{exp}(z)=\phi(0)\cdot\Psi\left(\frac{z}{3H_\mathrm{mono}}\right)\label{eq:exp_distr_profile}
\end{equation}
where $y=\Psi(x)$ is the inverse of the function 
$x=\sqrt{1-y}-\sqrt{y}\arcsin\left(\sqrt{1-y}\right)$.
The brush height is $H=3H_\mathrm{mono}$. \RE{This last result differs from the
value obtained by MC simulations $H\simeq 2H_\mathrm{mono}$ (see Fig.\ref{fig:brush_heights}).
This is not surprising, as the brush profile has a very long and flat tail at high polydispersities,
and the determination of the brush height from MC simulations becomes difficult. In other respect, the theoretical
predictions and the MC simulation data match reasonably well (see below).}

The full Green's function for a chain of arbitrary length $N$ is
not available. However, one can derive an exact expression for the average
end chain position as a function of $N$:
\begin{eqnarray}
Z_{e}(N)&=&\frac{6N_{n}}{\pi}\Big(\frac{2\phi_{0}}{3}\Big)^{1/2}\left[\left(1-\exp\left(-\frac{2N}{3N_{n}}\right)\right)^{1/2}\right.\nonumber\\
&-&\left.\exp\left(-\frac{2N}{3N_{n}}\right)\arcsin\left(1-\exp\left(-\frac{2N}{3N_{n}}\right)\right)^{1/2}\right]\nonumber\\
\end{eqnarray}

All other details are given in Appendix \ref{sec:theory-appendix}.

\section{Simulation results and comparison with theory}\label{sec:results}

We now discuss the results from MC simulations, analytical theory as well as
numerical SCF calculations in one dimension. The one dimensional SCF approach 
used in these calculations is outlined in Appendix \ref{SCF_1d_appendix}.

\subsection{Monodisperse and weakly polydisperse polymer brushes}
\label{sec:monodisperse}

\begin{figure}[t]
  \centering
  \subfigure[]{
    \label{fig3a} 
    \includegraphics[angle=0, width=6.0cm]{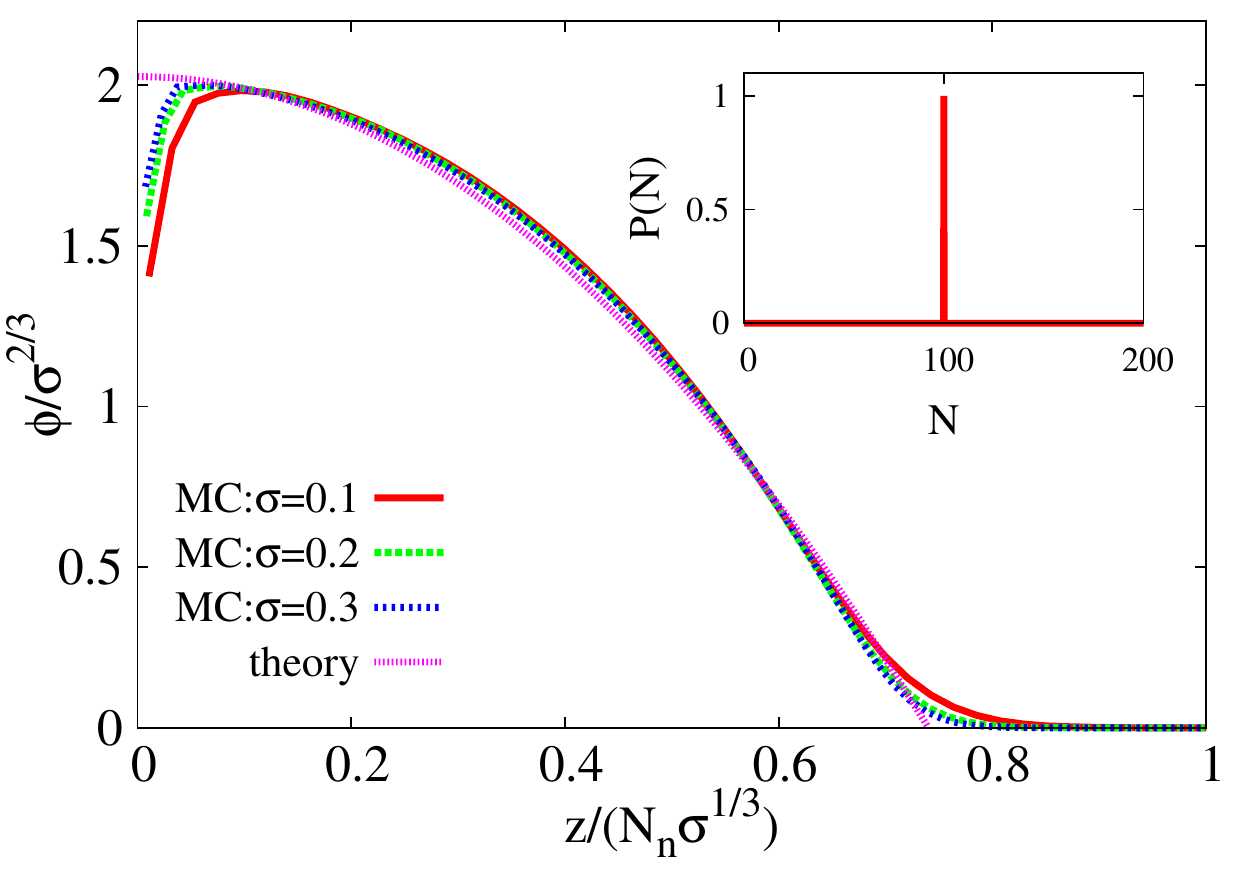}}
  \subfigure[]{
    \label{fig3b}
    \includegraphics[angle=0, width=6.0cm]{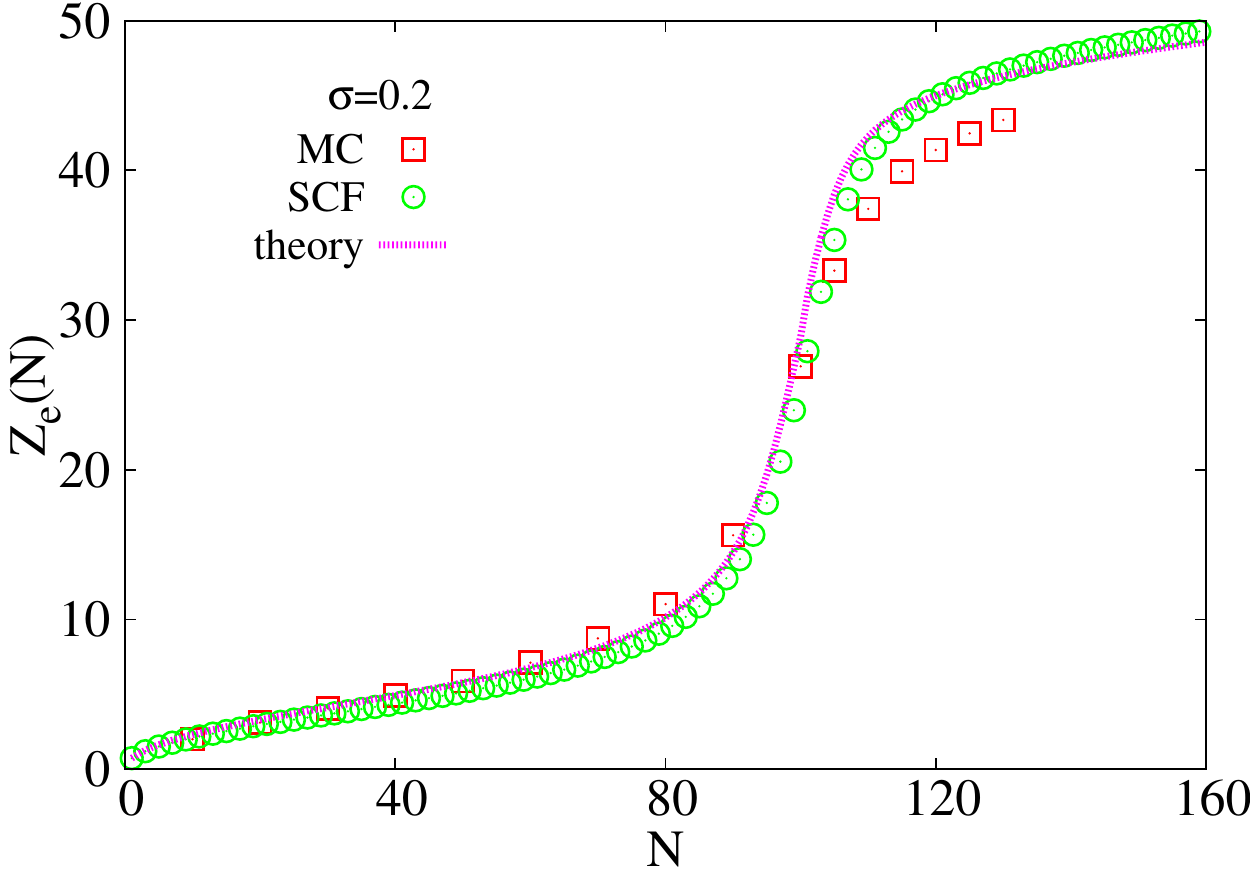}}
  \subfigure[]{
    \label{fig3c}
    \includegraphics[angle=0, width=6.0cm]{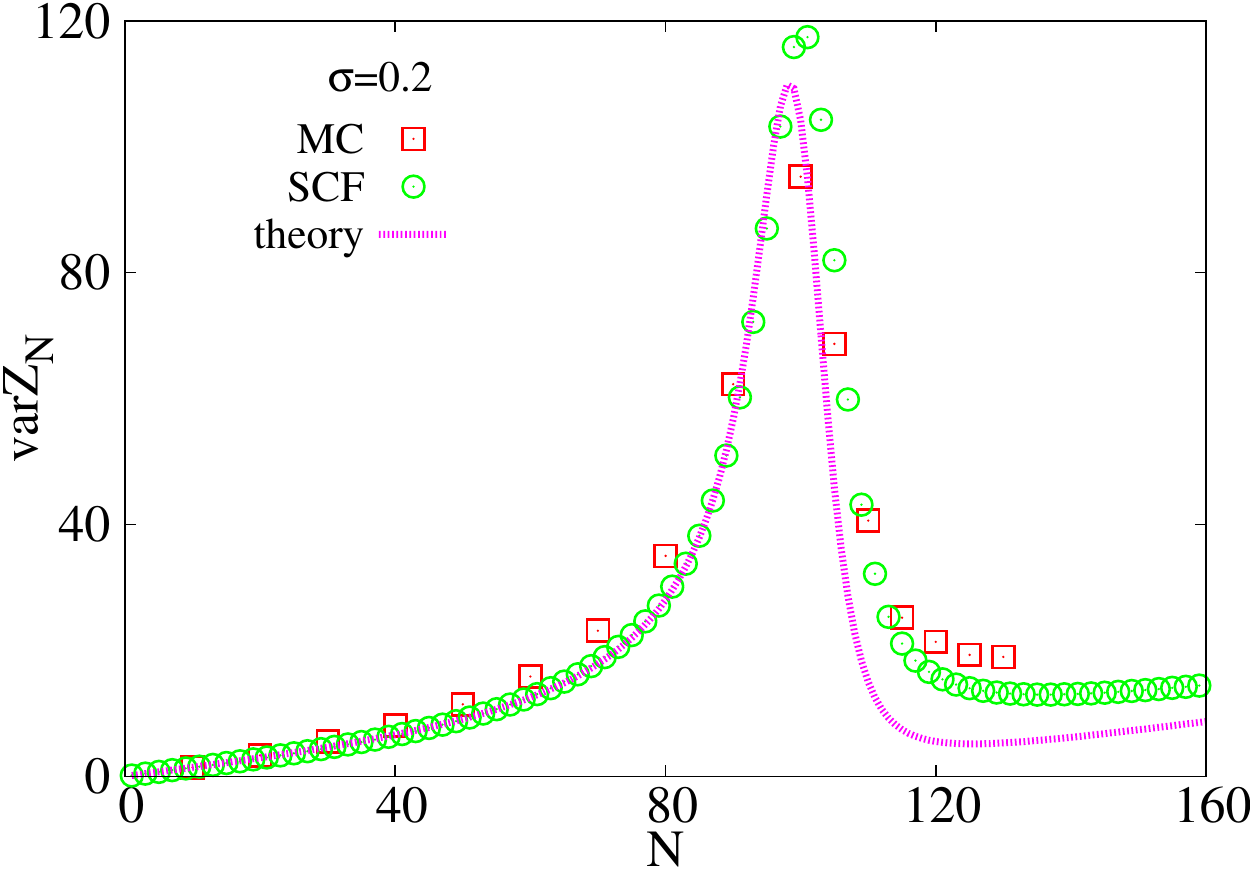}}
  \caption{(a): Density profiles from MC simulations in rescaled coordinates
$\phi(z)/\sigma^{2/3}$ vs $z/(N_n\sigma^{1/3})$
for monodisperse brushes at $N_n=100$ and grafting densities 
$\sigma=0.1,$ 0.2 and 0.3.
The chain length distribution $P(N)$ of the brush chains is a delta-peak, 
see inset.  Solid purple line corresponds to the 
parabolic density profile predicted by \eqn (\ref{eq:mono_profile}). 
(b): Average end position $Z_e(N)$ and (c): fluctuations of the end positions 
var$Z_N$ (c) of isolated minority chains of length $N$ inserted in the 
monodisperse brush at grafting density $\sigma=0.2$ as obtained from 
Monte Carlo simulations (red squares) and numerical self-consistent 
field calculations (green circles).
Solid purple lines shows the corresponding theoretical predictions
given in Appendix \ref{sec:theory_mono}, which are calculated from
\eqn (\ref{eq:Green_parabolic}). 
}
  \label{fig:monodisperse} 
\end{figure}

We first consider the monodisperse limit (\mbox{$k\to\infty$}, \mbox{$\pdi=1$})
as a reference. \fig \ref{fig:monodisperse} compares the MC simulations data
and the SCF calculations with the theoretical predictions for monodisperse
brushes at $N_n=100$ and grafting densities $\sigma=0.1,$ 0.2 and 0.3.  Panel
(a) displays the density profiles in rescaled coordinates
$\phi(z)/\sigma^{2/3}$ vs $z/(N_n\sigma^{1/3})$. The chain length distribution
has the form of a delta-function and is shown in inset. Panels (b) and (c) show
the average end position and fluctuations for minority chains inserted in the
otherwise monodisperse brush of chain length $N_n$. The length of the minority
chain is denoted as $N$ and it can be shorter or longer than $N_n.$

\begin{figure}[h]
  \centering
  \subfigure[]{
    \label{fig4a} 
    \includegraphics[angle=0, width=6.0cm]{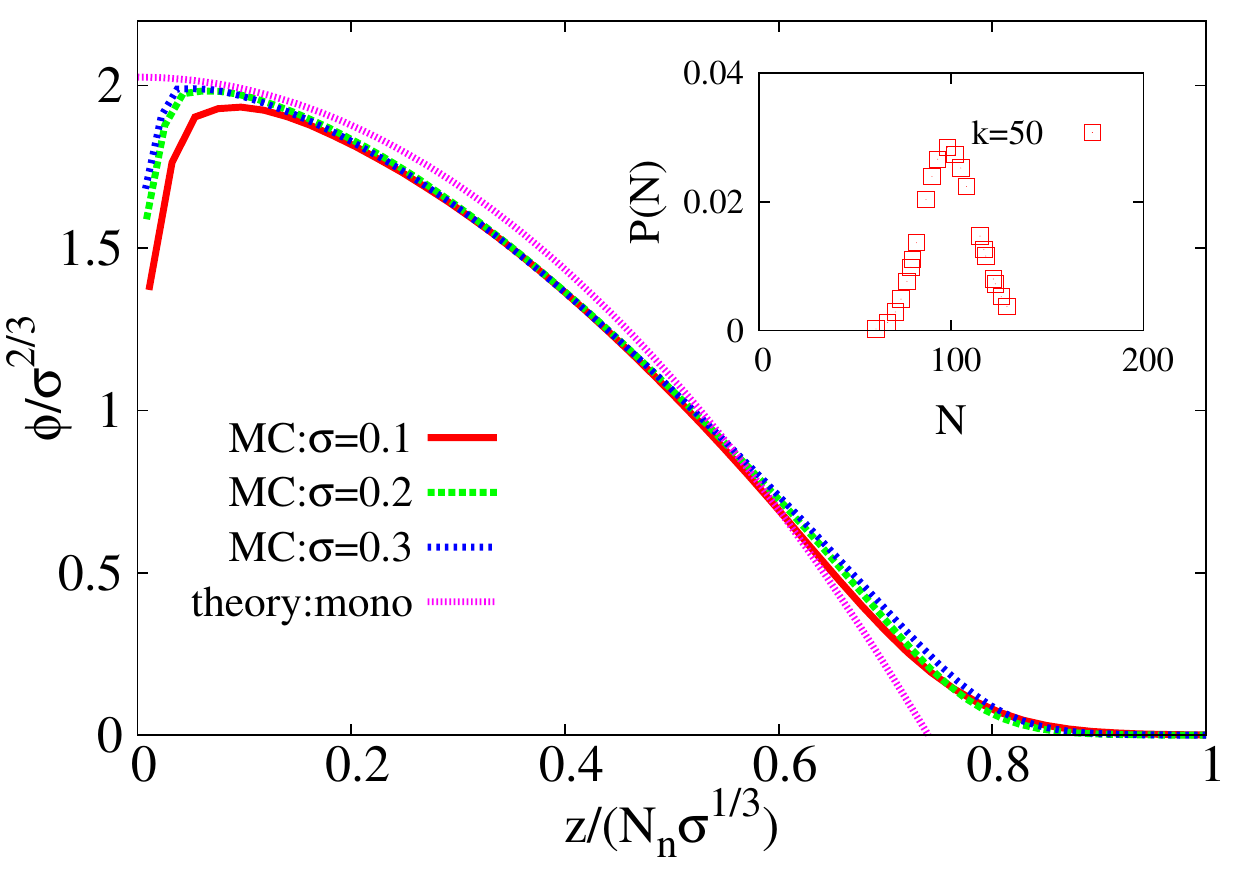}}
  \subfigure[]{
    \label{fig4b}
    \includegraphics[angle=0, width=6.0cm]{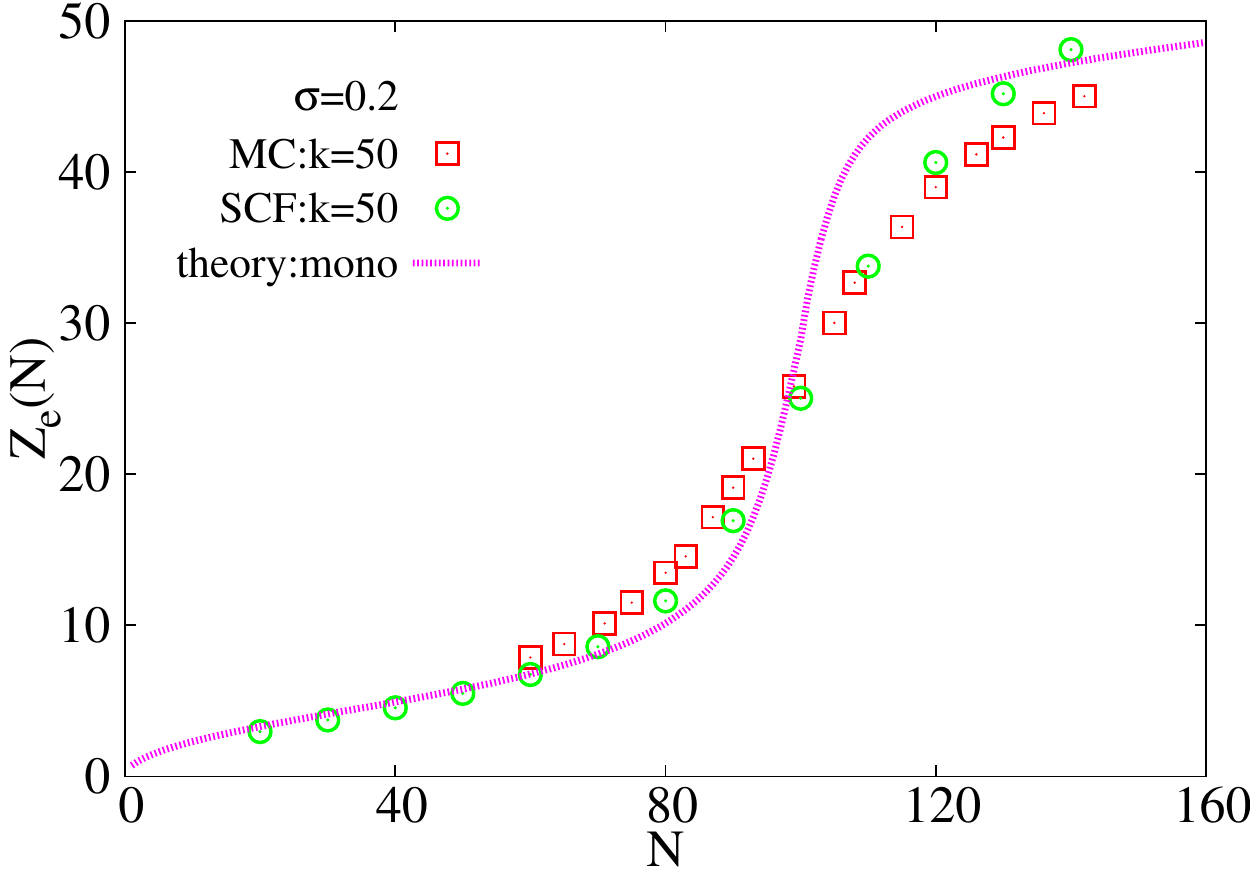}}
  \subfigure[]{
    \label{fig4c}
    \includegraphics[angle=0, width=6.0cm]{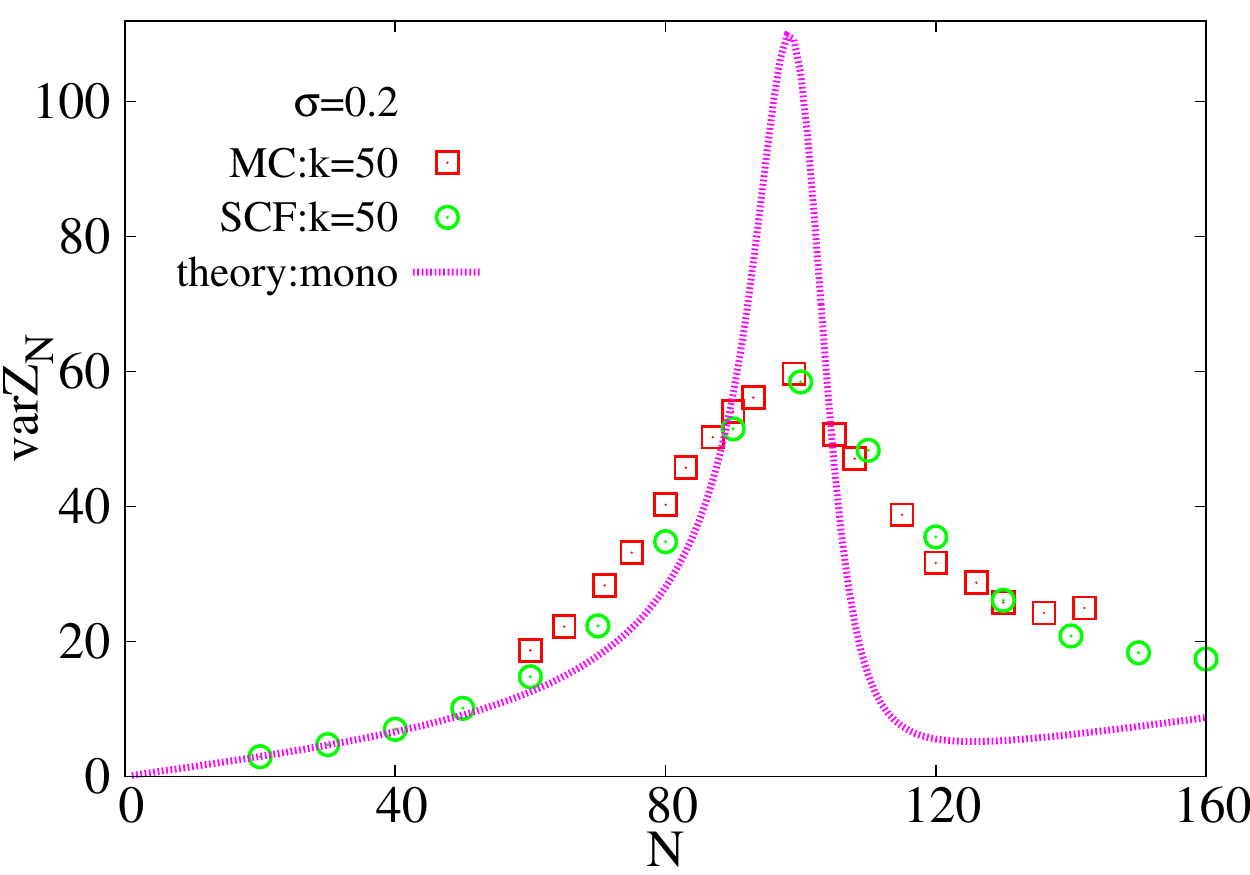}}
  \caption{Same as \fig \protect\ref{fig:monodisperse} for weakly
polydisperse brushes with Schulz-Zimm distribution at $k=50$
with $N_n=100$, $\pdi=1.02$. The corresponding chain length
distribution is shown in the inset of (a)). 
Solid purple lines show the theoretical predictions for monodisperse 
brushes (same curves as in \fig \protect\ref{fig:monodisperse}.}
  \label{fig:weakly_polydisperse} 
\end{figure}

To enable a comparison with polydisperse brushes, where the length of the
individual constituent chains generally differs from the number average $N_n$,
we add isolated minority chains with length $N \neq N_n$ to the monodisperse
brush and study their behavior.  A sharp increase in the end-height
fluctuations as $N\rightarrow N_{n}$, see panel (c), signals the onset of
critical-type behavior of individual chains in the monodisperse brush.  The
critical behavior of individual chains is inherently linked to monodispersity
and is suppressed in polydisperse brushes as will be discussed below
sec.\ref{sec:critical}.  Next we study a very weakly polydisperse brush with
$k=50$, $\pdi=1.02$.  Such a brush would be considered monodisperse in most
experimental setups. For such weak polydispersities with $k^{-1}\ll1$, the SZ
distribution is a narrow symmetric peak, see the inset of \fig
\ref{fig:weakly_polydisperse}(a). The density profile is still close to the
theoretical parabolic shape of \eqn (\ref{eq:mono_profile}). The main
difference is that the tenuous tail at the outer edge of the brush is more
strongly pronounced. However, the behavior of the end position $Z_e(N)$ for
chains of length $N$ is distinctly different from the truly monodisperse case.
The average end positions of the constituent chains are described by a
well-defined increasing function $Z_{e}(N)$, which is less sharp than that for
minority chains described in the previous subsection, see panel (b). The peak
of the fluctuations around $N=N_{n}$ is smaller and broader in comparison to
the monodisperse brush, see panel (c). This confirms the conclusions of Ref.\
\cite{Klushin:1992}, where anomalous fluctuations were predicted to be
suppressed already in weakly polydisperse brushes.  

\subsection{Moderately polydisperse brush}

\begin{figure}[t]
  \centering
  \subfigure[]{
    \label{fig6a}
    \includegraphics[angle=0, width=6.0cm]{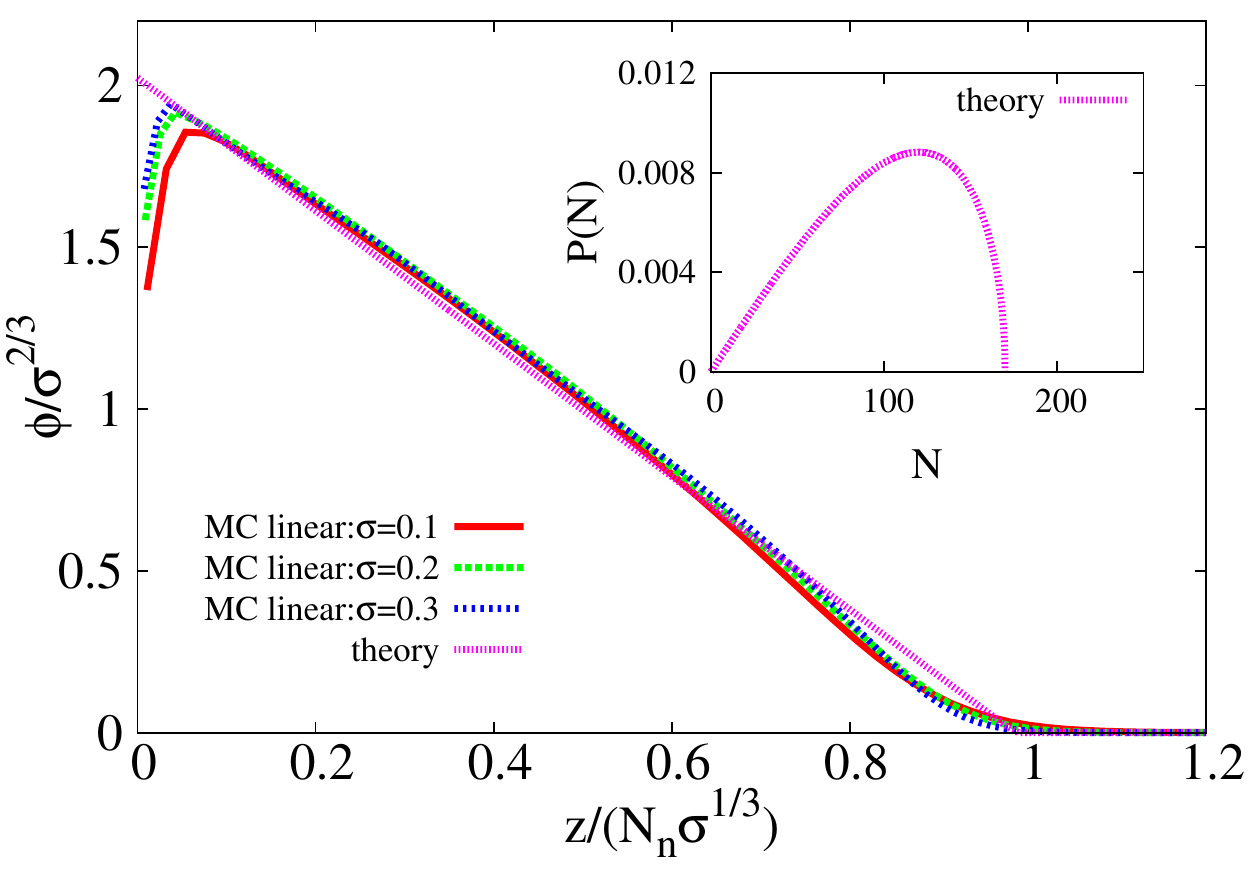}}
  \subfigure[]{
    \label{fig6b}
    \includegraphics[angle=0, width=6.0cm]{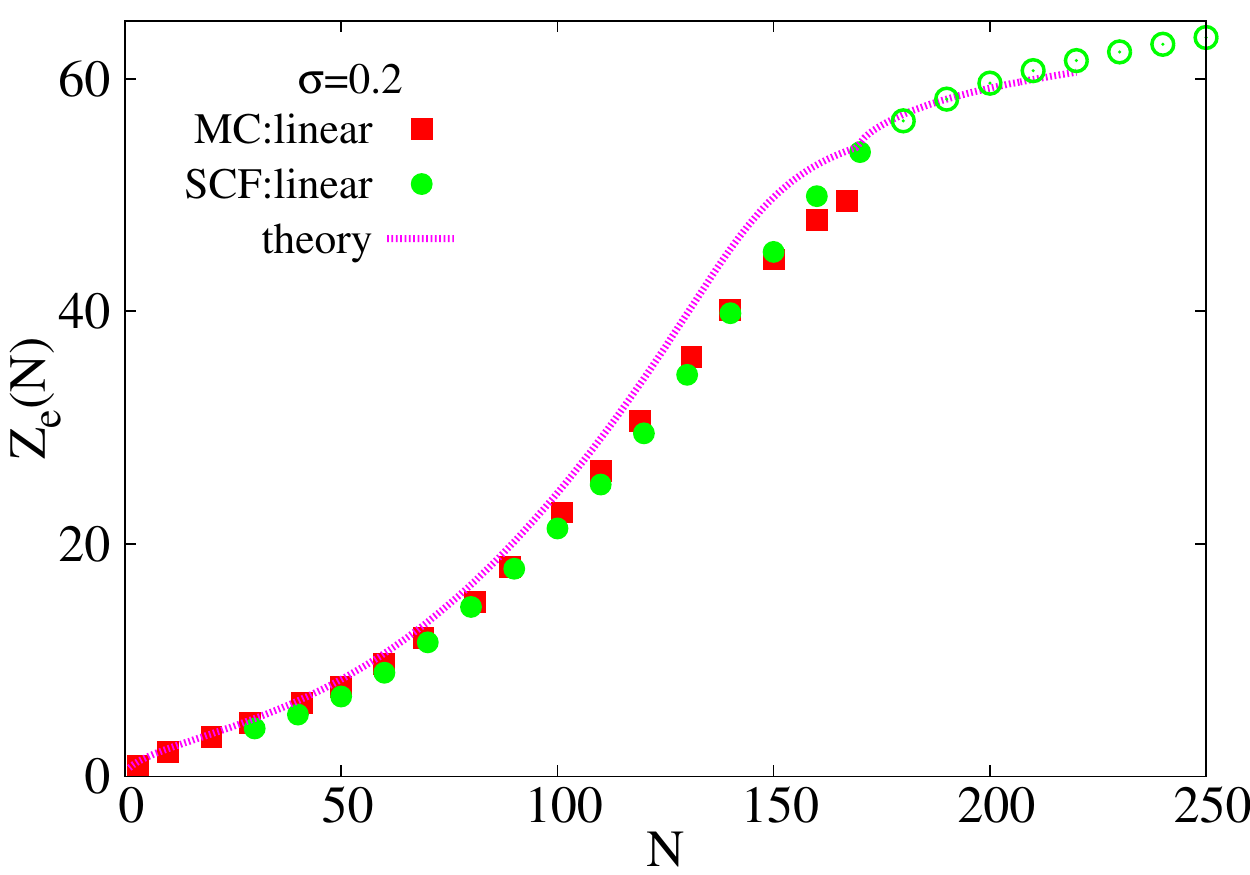}}
  \subfigure[]{
    \label{fig6c}
    \includegraphics[angle=0, width=6.0cm]{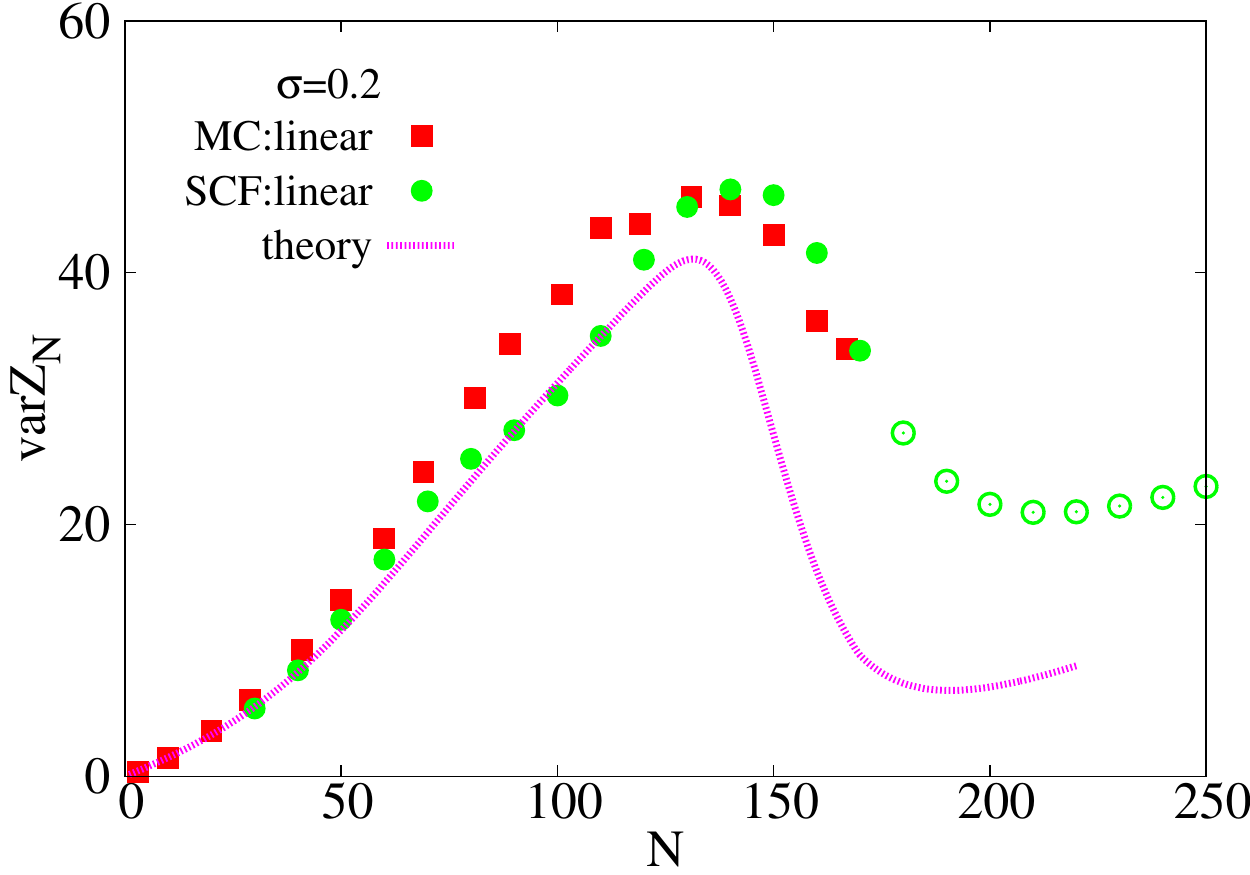}}
  \caption{Same as \fig \protect\ref{fig:monodisperse} for 
moderately polydisperse brushes with chain length distribution
given by \eqn (\ref{eq:CLD_linear_profile})
(shown in the inset of (a)) with $N_n=100$ and $\pdi=1.15$,
leading to a maximum chain length $N_\mathrm{max} = 170$.
Solid purple lines show theoretical predictions for brushes
with linear density profiles as derived from \eqn (\ref{eq:cut_off_effect})
in Sec.\ \protect\ref{sec:theory_linear}.}
  \label{fig:moderate_polydisperse_linear}
\end{figure}

We turn to discuss brushes with moderate polydispersity, focusing in particular
on the case discussed in Sec.\ \ref{sec:theory_linear}, where the brush density
profile is close to linear. For the SZ distribution, this is reached at $k=7$,
corresponding to $\pdi=1.14$. In Appendix
\ref{sec:theory_linear}, we also derive a chain length distribution
profile that generates a strictly linear density profile in the strong
segregation limit (\eqn (\ref{eq:CLD_linear_profile})). In the following, we
will study and compare the structure of polydisperse brushes with these two
chain length distributions, i.e., the SZ distribution at $k=7$ and the
distribution of \eqn (\ref{eq:CLD_linear_profile}). We begin with the latter
(\fig \ref{fig:moderate_polydisperse_linear}).

\begin{figure}[t]
  \centering
  \subfigure[]{
    \label{fig6d} 
    \includegraphics[angle=0, width=6.0cm]{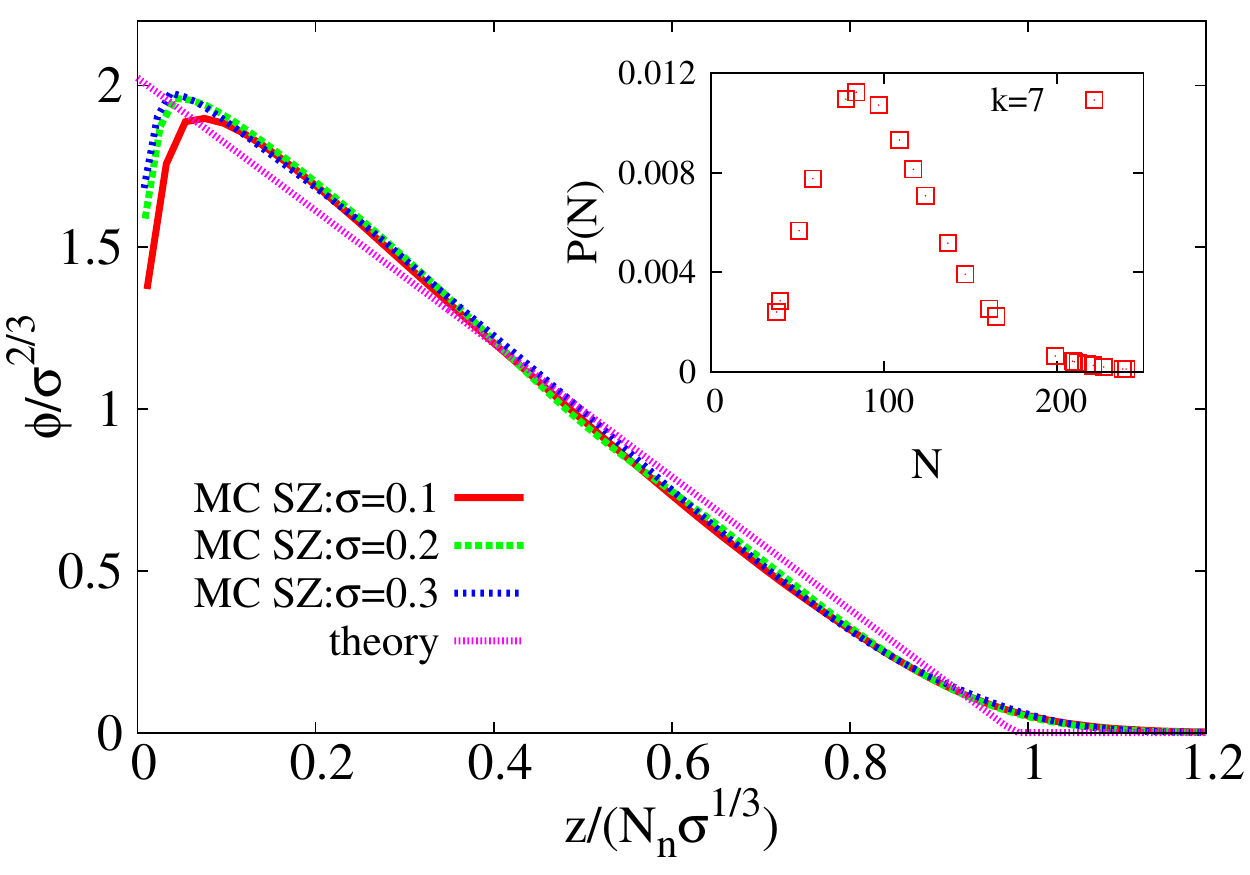}}
  \subfigure[]{
    \label{fig6e}
    \includegraphics[angle=0, width=6.0cm]{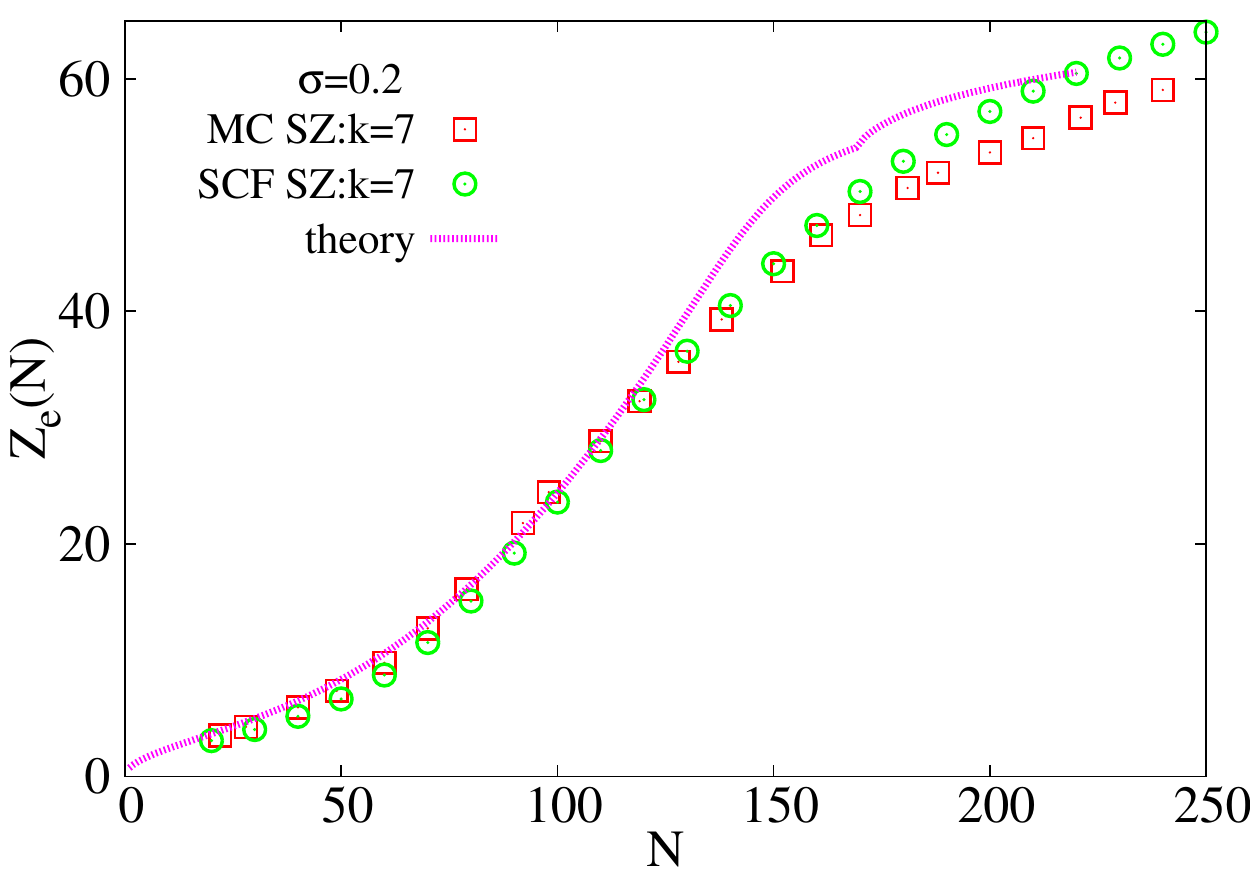}}
  \subfigure[]{
    \label{fig6f}
    \includegraphics[angle=0, width=6.0cm]{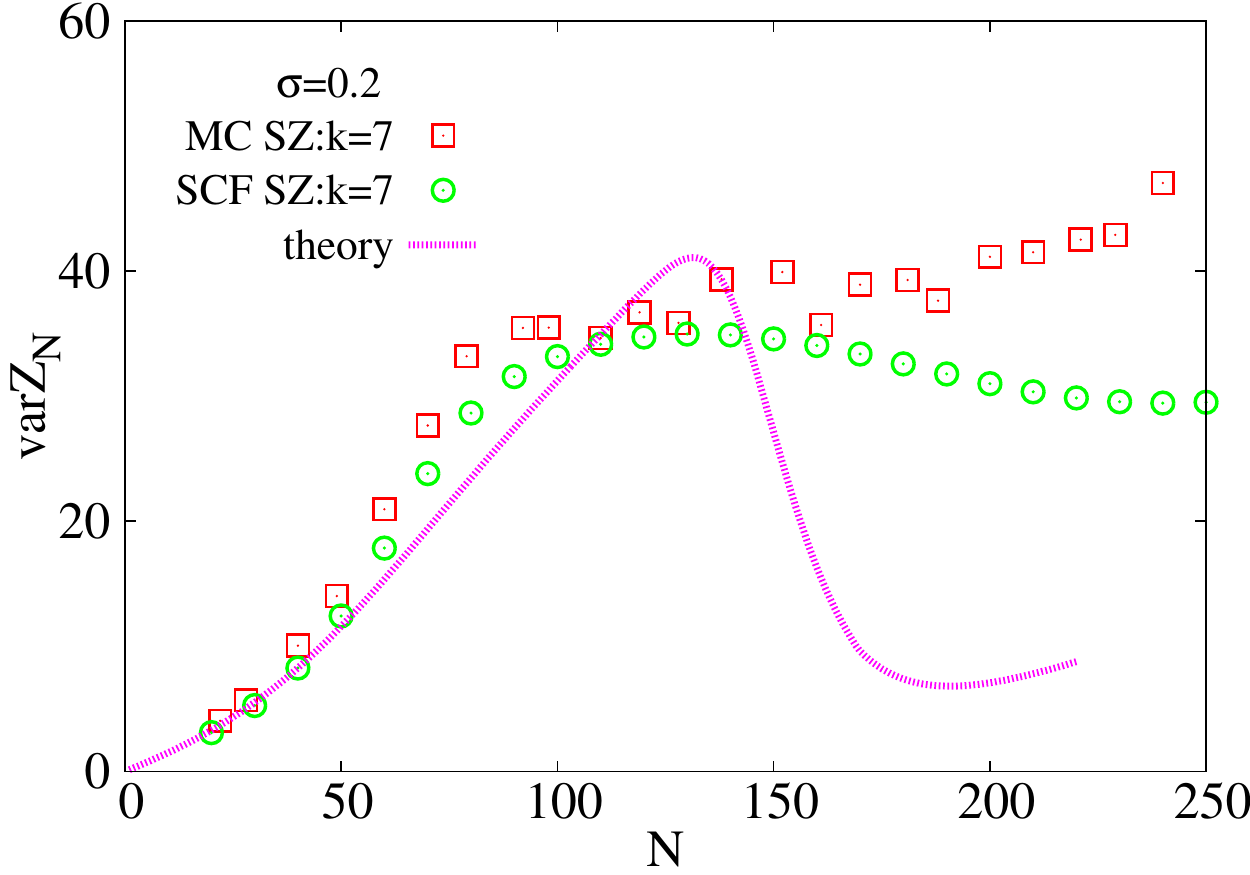}}
  \caption{
Same as \fig \protect\ref{fig:moderate_polydisperse_linear}
for moderately polydisperse brushes with SZ chain length distribution 
at $k=7$ with $N_n=100$ and $\pdi=1.14$.
Inset in panel (a) shows the corresponding chain length distribution.
Solid purple lines show the same theoretical curves than in 
\fig \protect\ref{fig:moderate_polydisperse_linear}.}
  \label{fig:moderate_polydisperse}
\end{figure}

\fig \ref{fig:moderate_polydisperse_linear}(a) demonstrates that the density
profile obtained by MC simulations is, indeed, extremely close to linear even
for moderately long chains with $N_{n}=100$.  Theoretical curves for the
average end positions and fluctuations are calculated numerically based on the
Green's function, \eqn (\ref{eq:Green_function_at_k7}).  Our specific chain
length distribution function has a the polydispersity index $\pdi=1.15$ and a
strict cut-off at the maximum chain length
$N_\mathrm{max}=\frac{16}{3\pi}N_{n}$.  In order to also assess chain end
positions and fluctuations for longer chains, we use the same procedure than
for the monodisperse chains and calculate the properties of longer probe chains
inserted in the brush with linear profile.  The application of the Green's
function formalism to chains that extend outside the brush edge is explained in
Appendix \ref{sec:theory_linear} at the example of monodisperse brushes.  

The theory predicts the mean positions of chain ends reasonably well 
(\fig \ref{fig:moderate_polydisperse_linear}(b));
in particular, the middle part of the curves follow a simple quadratic
dependence $Z_{e}(N)=N^{2}f/6$, as expected for a free chain in a gravitational
field, where $f$ is the force per monomer given by
\eqn (\ref{eq:force_in_linear_field}).  Close to the substrate and at the brush
surface, the profile deviates from the quadratic behavior due to boundary
effects. Short chains are affected by the hard substrate and have mushroom
conformations.  Long chains with length close to $N \approx N_\mathrm{max}$ or
longer do not feel the full ``gravitational'' force at the edges. 
The chain end fluctuations, shown in \fig
\ref{fig:moderate_polydisperse_linear}(c), exhibit a pronounced maximum
followed by a drop which is due to the boundary effect of the outer brush edge.
The theory provides a qualitatively correct prediction of the maximum and the
subsequent drop for chains with length $N\simeq N_{max}$ and slightly longer. 

\fig \ref{fig:moderate_polydisperse} shows the corresponding data for brushes
with SZ distribution at $k=7$ ($\pdi=1.14$). The behavior is mostly similar to
that shown in \fig \ref{fig:moderate_polydisperse_linear}.  The density has a
smoother profile at the outer edge. This leads to a significant difference in
the fluctuation curve where the non-monotonic behavior degenerates into a
plateau.

\subsection{Strongly polydisperse brushes}

\begin{figure}[t]
  \centering
  \subfigure[]{
    \label{fig7a} 
    \includegraphics[angle=0, width=6.0cm]{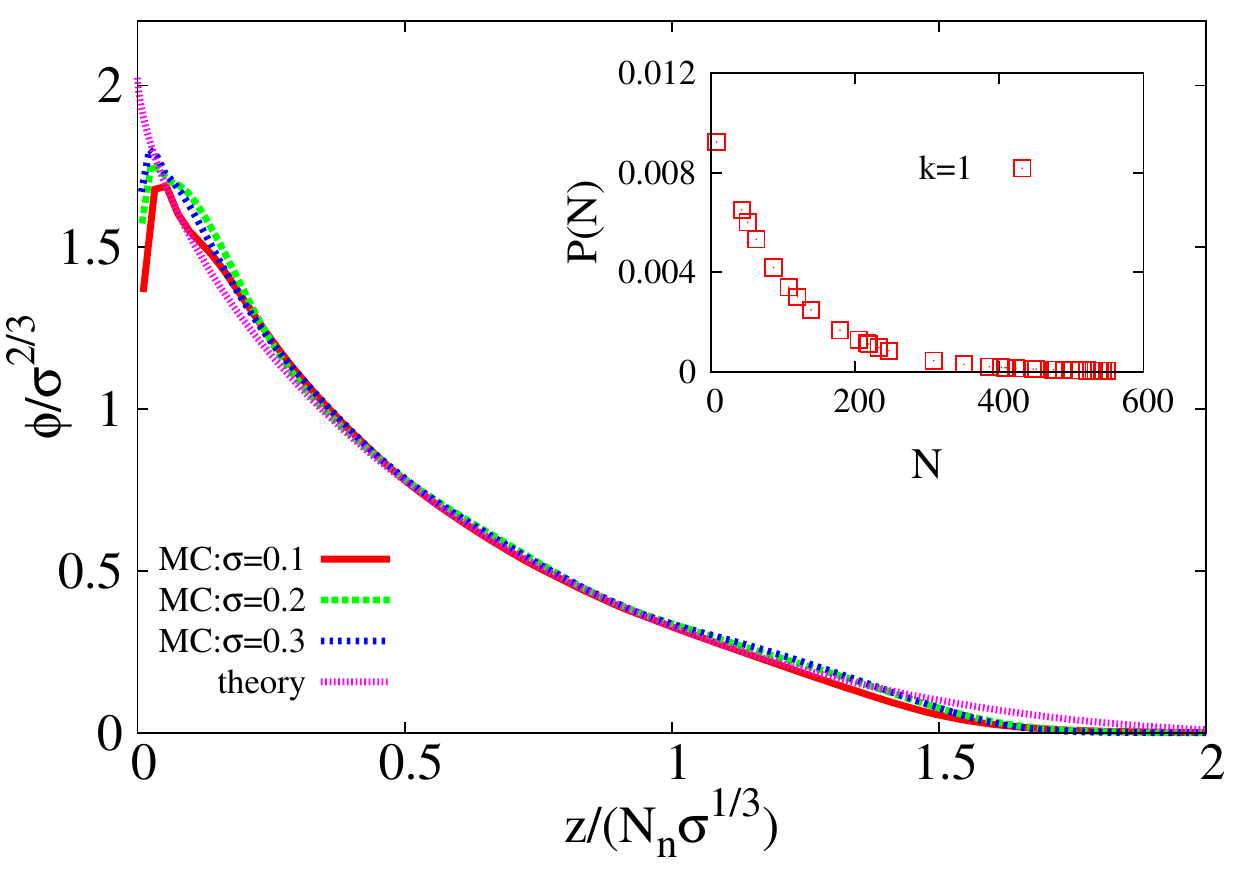}}
  \subfigure[]{
    \label{fig7b}
    \includegraphics[angle=0, width=6.0cm]{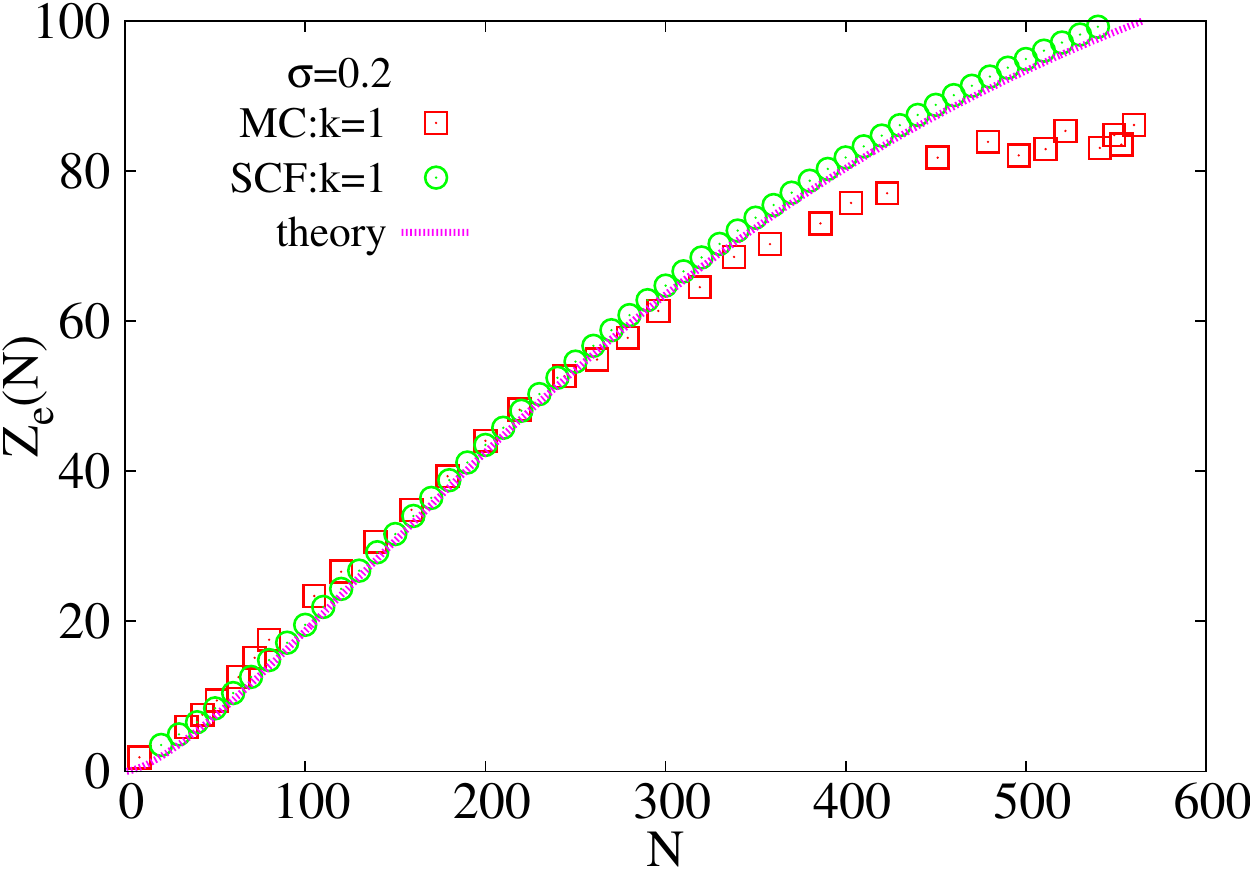}}
  \subfigure[]{
    \label{fig7c}
    \includegraphics[angle=0, width=6.0cm]{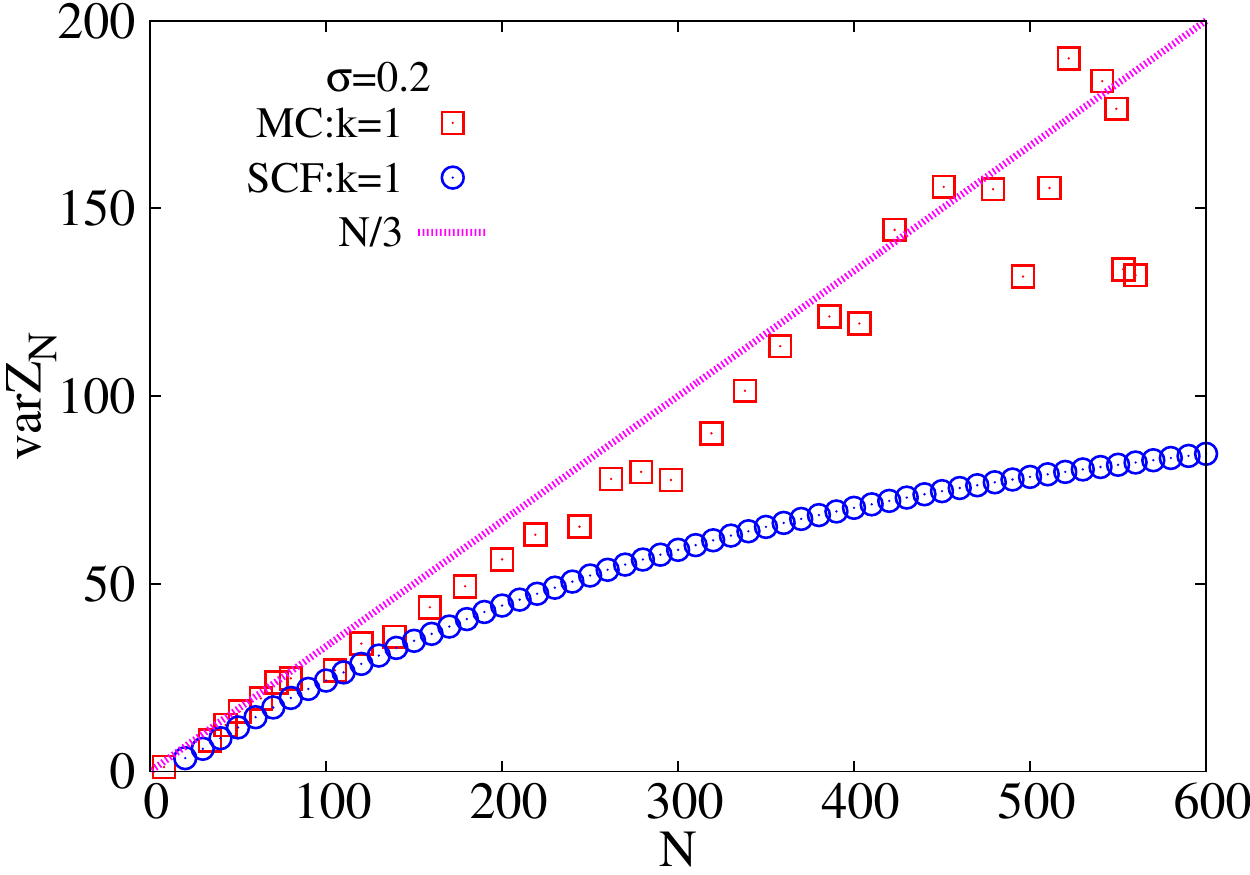}}
  \caption{
Same as \fig \protect\ref{fig:moderate_polydisperse_linear}
for strongly polydisperse brushes with SZ chain length distribution at
$k=1$ with $N_n=100$ and $\pdi=2$.
Inset in panel (a) shows the corresponding chain length distribution.
Solid purple line in (a) shows the theoretical density profile from
\eqn (\ref{eq:exp_distr_profile}), solid line in (b) the chain end 
distribution obtained from \eqns (\ref{eq:UN_exp}) and (\ref{eq:zU_exp}),
and solid line in (c) the end fluctuations of an ideal coil.}
  \label{fig:strongly_polydisperse} 
\end{figure}

Finally, we consider a strongly polydisperse brush with SZ distribution
characterized by the polydispersity parameter $k=1$ and, correspondingly, a
polydispersity index of $\pdi=2$. This case is discussed theoretically in
Appendix \ref{sec:theory_exp}. The brush density profile has a pronounced
convexity, see panel (a) of \fig \ref{fig:strongly_polydisperse}, while the
chain length distribution is reduced to a simple exponential (inset in panel
(a)). The analytical theory describes the \RE{density profiles from MC simulations
reasonably well} except for the
small depletion zone near the wall (which cannot be accounted for in the
strong-stretching approximation), and details at the outer brush edge which are
due to numerical difficulties of accurately sampling the tail of the chain
length distribution in the MC simulation.  Panel (b) displays the average end
monomer positions as a function of the length of constituent chains. Again,
there is a slight discrepancy between the MC data and the theory in the range
of the longest chains, which is most probably related to the corresponding
difference in the density profiles, while the agreement of the SCF results and
the theory is excellent.  The chain end fluctuations are presented in \fig
\ref{fig:strongly_polydisperse}(c). The analytical theory cannot describe the
fluctuations since it is based on a Newtonian-path approximation. Surprisingly,
the MC simulation results can be fitted by the simple formula for the ideal
coil fluctuations: var$Z_{N}=\frac{N}{3}$. SCF calculations predict a
suppression of end fluctuations with respect to the ideal coil values, which
can be related to the convex shape of the effective potential. However, this is
not observed in the simulations.  In mean-field theory, chains are subject to a
weak effective potential generated by other chains for all distances from the
substrate. In practice, however, the longest chains are fairly isolated within
the brush, and mostly behave as free coils.

\subsection{Monodisperse brushes as near-critical systems }

\label{sec:critical}

\begin{figure}[t]
\centerline{\includegraphics[width=7cm]{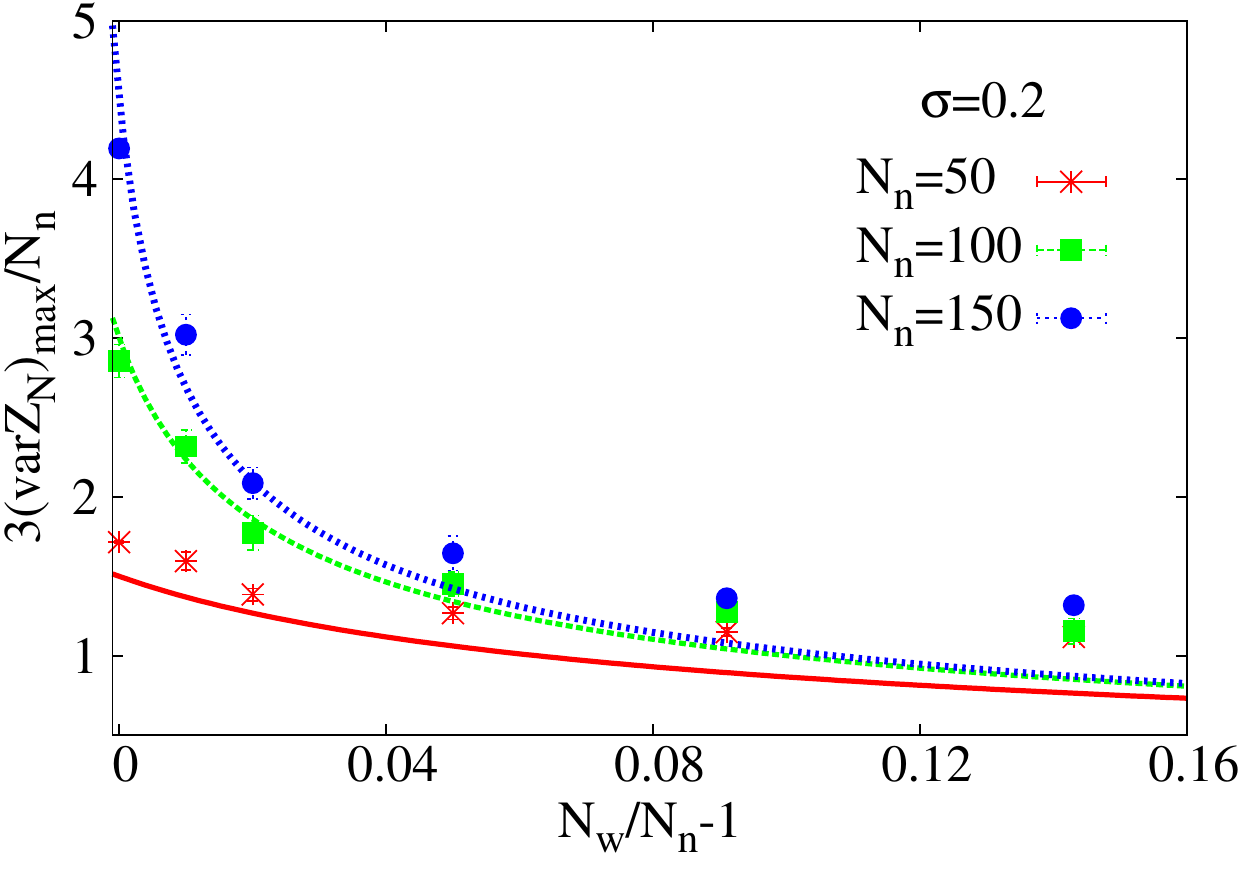}}
\centerline{\includegraphics[width=7cm]{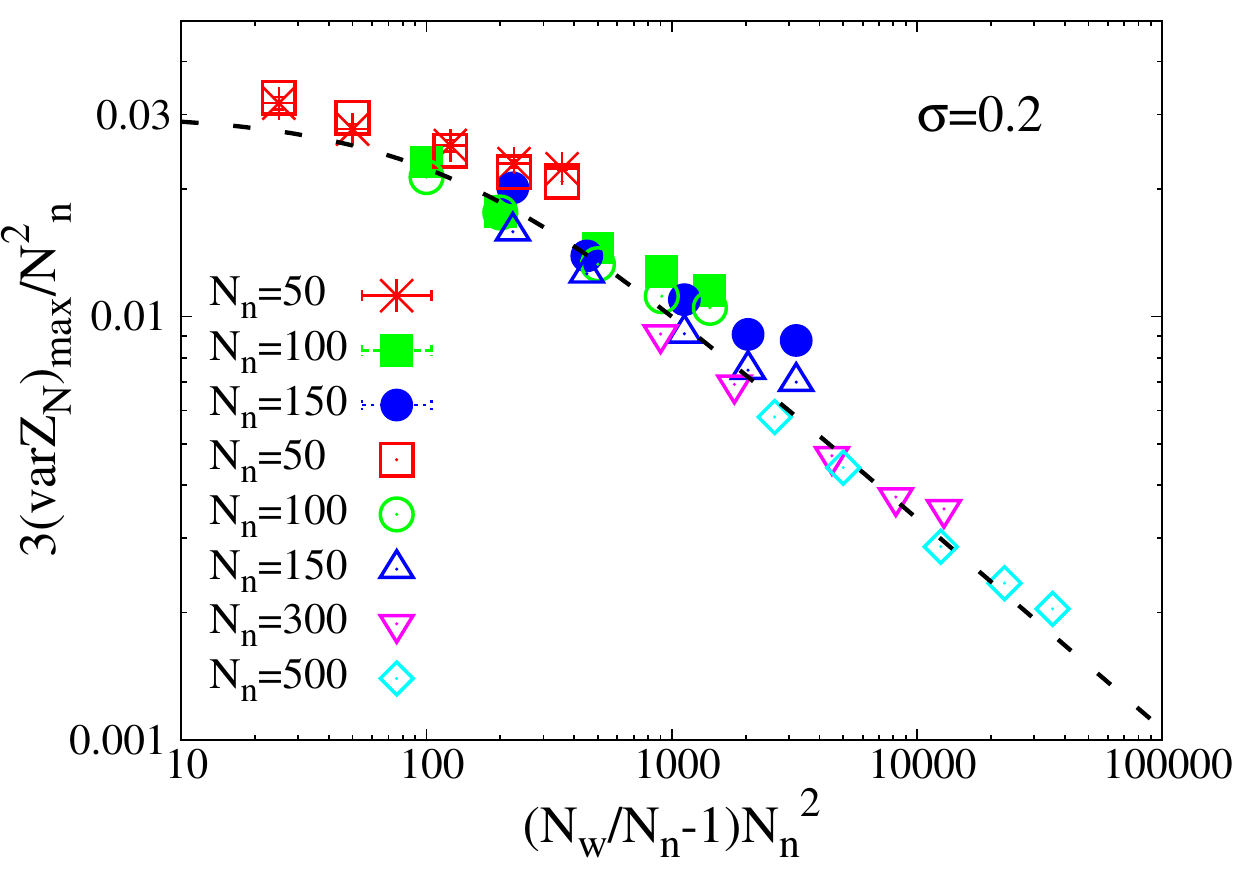}}
\caption{
a) Maximal enhancement of chain end fluctuations in brushes
compared to ideal coils as a function of the polydispersidy index $\pdi$.
b) Rescaled plot according to \eqn (\protect\ref{eq:enhancement_scaling})
(with $\delta_1 \sim N_n^{-1}$ and $\delta_2 \sim (\pdi -1)$), 
showing that the data roughly collapse onto one master curve.
\RE{The filled symbols correspond to the results from MC simulations, 
the empty symbols are data from SCF calculations.
The dashed line in b) shows an empirical scaling
function of the form $\hat{V}(x)=\hat{V}(0)/\sqrt{1+b x}$, where
$\hat{V}(0)=0.03$ is obtained from Eq.\ (\protect\ref{eq:mono_end_fluct})
(the theoretical prediction for monodisperse brushes), and $b=0.008$ is
an adjustable parameter. Lines in a) give the results for the same scaling 
function in the original variables.}
}
\label{fig:The-maximal-fluctuations} 
\end{figure}

From the discussion above, it is clear that polydispersity has a pronounced
effect on the chain end fluctuations. \fig
\ref{fig:The-maximal-fluctuations}(a) displays the peak values of the chain end
fluctuations as a function of the polydispersity index $\pdi$.  For small
polydispersities, the peak of $\mathrm{var}Z_{N}(N)$ is localized near the
point $N=N_{n}$ (see \fig\ref{fig4c}). At higher polydispersities, the peak first degenerates into a
plateau in the $N\sim N_{n}$ range  (see \fig\ref{fig6f}), and then disappears (see \fig\ref{fig7c}).  Both theory and simulations indicate that at
sufficiently high polydispersity, the chain end fluctuations are essentially
reduced to the values characteristic of an ideal coil, $\mathrm{var}Z_{N}=N/3$.
Therefore, the peak values displayed in the \fig
\ref{fig:The-maximal-fluctuations}(a) have been normalized by
$N_{n}/3$, thus demonstrating the enhancement of fluctuations compared to the
ideal coil. With increasing polydispersity, the enhancement factor initially
drops sharply and then decays further and approaches 1.

A monodisperse brush
exhibits anomalous chain fluctuations \cite{Klushin:1991}, see Eq.(\ref{eq:mono_end_fluct}). The enhancement factor is given by
$3{\mathrm{var}Z_\mathrm{mono}}/{N_n}\sim\sigma^{2/3}N_n$ and diverges for
asymptotically long chains. 
Polydispersity clearly serves as a parameter that suppresses the anomalous
fluctuations enhancement.  One is naturally reminded of the famous
polymer-magnetic analogy proposed by des Cloizeaux and Jannink
\cite{Polymer_solution}. \RE{The phase behavior of a ferromagnet close to the
critical point is driven by two relevant scaling fields, the temperature and the
external magnetic field. In a similar manner, we can also identify two parameters,
the number-averaged chain length and the polydispersity parameter, which
determine the critical behavior of the polydisperse brushes.
}

In fact an isolated polymer coil by itself exhibits
near-critical density fluctuations characterized by a large correlation radius
$\xi\sim N^{\nu}$ (where $\nu$ is the Flory exponent). The inverse chain
length, $N^{-1}$, plays the role of the distance from a critical point. In a
semi-dilute solution long-range correlations are suppressed at the length of
$\xi\sim c^{-\frac{\nu}{3\nu-1}}$. The true critical point is recovered in the limit of
$N^{-1}\rightarrow0,\; c\rightarrow0$. For a monodisperse brush, the combination
$\delta_{1}\sim\sigma^{-2/3}N_{n}^{-1}$ serves as a measure of a distance from
a true critical point, the fluctuation enhancement factor diverging as
$\delta_{1}^{-1}$. It is worth noting that the parameter $\delta_1$ can be
associated with the (squared) ratio $\delta_1 \sim (R_g/H_{\mathrm{mono}})^2$
of two characteristic length scales of the system, namely the brush height
$H_{\mathrm{mono}}$ and the gyration radius $R_g \sim \sqrt{N}$ of ideal coils
of length $N_n$. Hence $\delta_1^{-1/2}$ can also be interpreted as a
stretching parameter, which diverges as $N_n$ becomes infinitely large at fixed
$\sigma$, and $\delta_1^{-1}$ is proportional to the average energy required to
insert a chain into the brush.

The polydispersity parameter, $(\pdi-1)$ is
linked to yet another distance, $\delta_{2}$, from the critical point. Two of
us have shown \cite{Klushin:1992} that for a specific type of asymmetric narrow
chain length distributions, the fluctuation enhancement factor scales as
$(\pdi-1)^{-1/2} = \delta_2^{-1/2}$.  The origin of the polydispersity effect
on the chain fluctuations lies in the special property of the mean force
potential in a purely monodisperse brush. It is known that a broad chain end
distribution (with width of order $H$) is a necessary requirement for the
existence of a monodisperse brush in the limit of asymptotically long chains,
irrespective of the particular brush regime (solvent quality, chain stiffness,
etc.) \cite{Lai_Zhulina_1992}.  Such a broad distribution requires a
cancellation of the elastic forces in the chain at the level of the softest
Rouse mode. The parabolic potential of mean force serves exactly to this
effect. It is essential that the second derivative $\frac{d^{2}\omega}{dz^{2}}$
is a constant and takes the a critical value of
$3\big(\frac{\pi}{2N_{n}}\big)^{2}$ (in absolute numbers). Even a very minor
deviation from perfect monodispersity changes the potential shape and leads to
a decrease in $\frac{d^{2}\omega}{dz^{2}}$.  Hence, the chain elasticity is not
cancelled completely, and this results in a dramatic reduction in the chain end
fluctuations. In Appendix \ref{sec:theory_field_weakly_p}, we demonstrate this
effect using a simple generic form of a chain length distribution linked to the
Fermi-Dirac distribution at low temperatures. This also results in a scaling of
the fluctuation enhancement factor of the form
\begin{equation}
\label{eq:enhancement_poly}
3\frac{[\mathrm{var}Z_\mathrm{N}]_\mathrm{max}}{N_n}\sim (\pdi -1)^{-1/2} 
  \sim \delta_2^{-1/2},
\end{equation}
independent of $\delta_1$. This scaling law is expected to hold in the limit
$\delta_1 \to 0$ or more generally in cases where $\delta_1$ is sufficiently 
small that the fluctuation enhancement of \eqn (\ref{eq:enhancement_poly}) 
is much smaller than that in the corresponding monodisperse brush.

In an attempt to unify the two different effects we make a scaling Ansatz
for the fluctuation enhancement factor 
$3(\mathrm{var}Z_{\mathrm N})_\mathrm{max}/N_{n} =: V(\delta_1, \delta_2)$,
which must be chosen such that we recover $V(\delta_1,0) \sim \delta_1^{-1}$
for monodisperse brushes, and $V(0,\delta_2) \sim \delta_2^{-1/2}$ for
infinitely long polydisperse brushes. These requirements are met
by the following Ansatz:
\begin{equation}
\label{eq:enhancement_scaling}
3\frac{(\mathrm{var}Z_\mathrm{N})_\mathrm{max}}{N_{n}}=
\delta_1^{-1} \;  \hat{V}(\delta_2 \; \delta_1^{-2}),
\end{equation}
where the scaling function $\hat{V}(x)$ approaches a constant for $x \to 0$ and
scales as $\hat{V}(x) \sim x^{-1/2}$ for large $x$.  To test this scaling
hypothesis, we replot in \fig \ref{fig:The-maximal-fluctuations}(b) the same
data as in \fig \ref{fig:The-maximal-fluctuations}(a) in a scaled
representation, i.e., we show the rescaled enhancement factor
$3(\mathrm{var}Z_\mathrm{N})_\mathrm{max}/{N_{n}} \: \delta_1 \sim
(\mathrm{var}Z_\mathrm{N})_\mathrm{max}/N_{n}^2$ as a function of the
scaling variable $\delta_2 \delta_1^{-2} \sim (\pdi-1) N_n^2$. \RE{In 
addition, \fig \ref{fig:The-maximal-fluctuations}(b) also includes 
SCF data for chain lengths up to $N_n=500$. The data for different average 
chain lengths $N_n$ roughly collapse. Deviation from the proposed 
scaling behavior can be noticed especially for short chains.
However, the asymptotic behavior 
$3(\mathrm{var}Z_\mathrm{N})_\mathrm{max}/N^2_n
  \propto [(N_w/N_n-1)N_n^2]^{-1/2}$ for long polydisperse chains
can be confirmed.
We should note that the critical exponents entering
\eqn(\ref{eq:enhancement_scaling}) are
mean-field exponents, in reality they may deviate.}
Nevertheless, we can conclude that a brush is a near-critical system as far as
the chain fluctuations are concerned. In order to observe this anomalous
behavior one needs to be close to the critical point characterized by the limit
$\left(N\sigma^{2/3}\right)^{-1}\rightarrow0,\; \pdi \rightarrow1$.

It is worth restating the consequences of this finding for situations where the
critical point is approached along the two directions in the parameter space. A
purely monodisperse brush always demonstrates near-critical chain fluctuations,
the relevant distance to the critical point being
$\delta_{1}\sim\sigma^{-2/3}N^{-1}$.  This results in slow (unentangled)
relaxation with the characteristic relaxation time scaling as $\tau\thicksim
N_{n}^{3}$ which was predicted theoretically \cite{Klushin:1991} and
subsequently verified by several simulations
\cite{Lai:1991,Lai:1994,Marko:1993,Virnau:2012}. The fluctuations enhancement
factor also describes the increase in the relaxation time compared to the Rouse
time, $\tau\simeq\tau_{R}\delta_{1}^{-1}$.  Another consequence of the
near-critical behavior is the increased chain susceptibility
\cite{Klushin:2014, Klushin:2015} leading to effects like the ``surface
instability'' introduced by Sommer and coworkers in Refs.\
\cite{Merlitz:2008,Romeis:2013, Romeis:2015}. There, the end-groups of some
brush chains were modified to become sensitive to small variations in the
solvent quality, which resulted in the possibility of an abrupt transition for
the modified chains either ``hiding'' inside the brush or exposing the
end-group at the outer brush edge. We should note that apart from the
individual chain fluctuations, a monodisperse brush exhibits ``normal''
behavior: for example, the density correlation length scales as
$\sigma^{-1/2}$, i.e. as the mean distance between the grafting points
\cite{Zhulina}.

Any finite polydispersity automatically limits the fluctuation enhancement.
The MC simulations data show that even at the polydispersity index of
$\pdi$=1.02, which is so close to monodisperse that it is actually quite
difficult to achieve experimentally, the near-critical behavior is severely
suppressed.  In the experimentally relevant polydispersity range
$\pdi=1.1\sim1.15$ the enhancement factor is of order 1. Thus we expect Rouse
relaxation times $\tau\sim N_{n}^{2}$ to be restored even in moderately
polydisperse brushes. In this range of polydispersities we also expect the
surface instability phenomenon to be considerably weakened if not to disappear
completely.

Although the limit $\pdi\rightarrow1$ of vanishing polydispersity is not very
interesting from a practical point of view, the fact that it is related to
critical behavior with a power-law fluctuation growth makes it quite intriguing
from the point of view of fundamental physics. Traditionally, this type of
behavior is associated with a continuous (second order) phase transition. Can
we identify the actual phase transition in brushes at vanishing polydispersity?
It is clear from the simulation results that the brush density profile
essentially does not change in the $\pdi\rightarrow1$ limit. Thus all the bulk
properties of the brush remain the same. It is difficult to identify an
extensive ``order parameter'' that would characterize a continuous change into
a new phase with a different symmetry, as in the magnetic counterpart of the
des Cloizeaux analogy. Qualitatively, the situation rather resembles a
supercritical fluid approaching the gas-liquid critical point, where the
density remains essentially constant and the symmetry of the underlying phase
does not change. However, the critical point can only be approached from one
side. The regime of two phases (gas-liquid) coexistence has no obvious
counterpart in the brush case. On the other hand, recent numerical SCF
calculations of Romeis and Sommer \cite{Romeis:2015} have revealed a possibly
related multicritical behavior in binary brushes where the two components $a,
b$ have different chain lengths $N_{a,b}$ and different solvent
selectivities $v_{a,b}$. These systems exhibit a multicritical point at
$N_a=N_b, v_a=v_b$, which connects two coexistence regions, one in the quadrant
$v_a<v_b, N_a > N_b$ and one in the quadrant $v_a > v_b, N_a < N_b$, where a
state with exposed $a$-chains coexists with a state with exposed $b$-chains.
It might also be possible to identify coexistence regions in polydisperse
brushes if we extend the space of control variables.

\section{Polydisperse Brush in Hydrodynamic Shear Flow}\label{sec:hydrodynamics}

Finally in this section, we discuss the interaction of a brush-coated 
surface with a hydrodynamic flow in the limit where the flow is so small
that its effect on the brush can be neglected \cite{Doyle:1998}. 
In general, the response of the brush to flow is controlled by the Weissenberg
number, $W_{i}$ , which is defined as the product of the brush chain relaxation
time and the typical shearing rate $W_{i}=\tau\dot{\gamma}$. Here we consider
the regime of vanishingly small $W_i$. A detailed understanding of
flow-mediated forces for higher Weissenberg numbers is much more challenging
and requires either approaches based on system-specific theories for the
hydrodynamics of complex fluids or explicit dynamic simulations.

\begin{figure}[t]
  \centering
  \subfigure[]{
    \label{fig9a} 
    \includegraphics[angle=0, width=7.0cm]{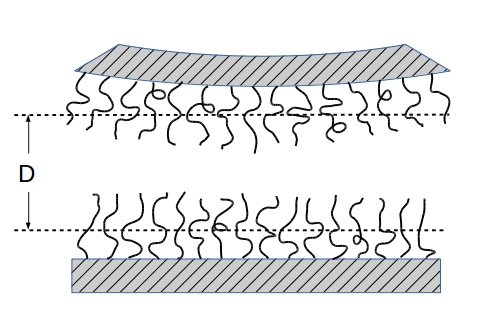}}
  \subfigure[]{
    \label{fig9b}
    \includegraphics[angle=0, width=7.0cm]{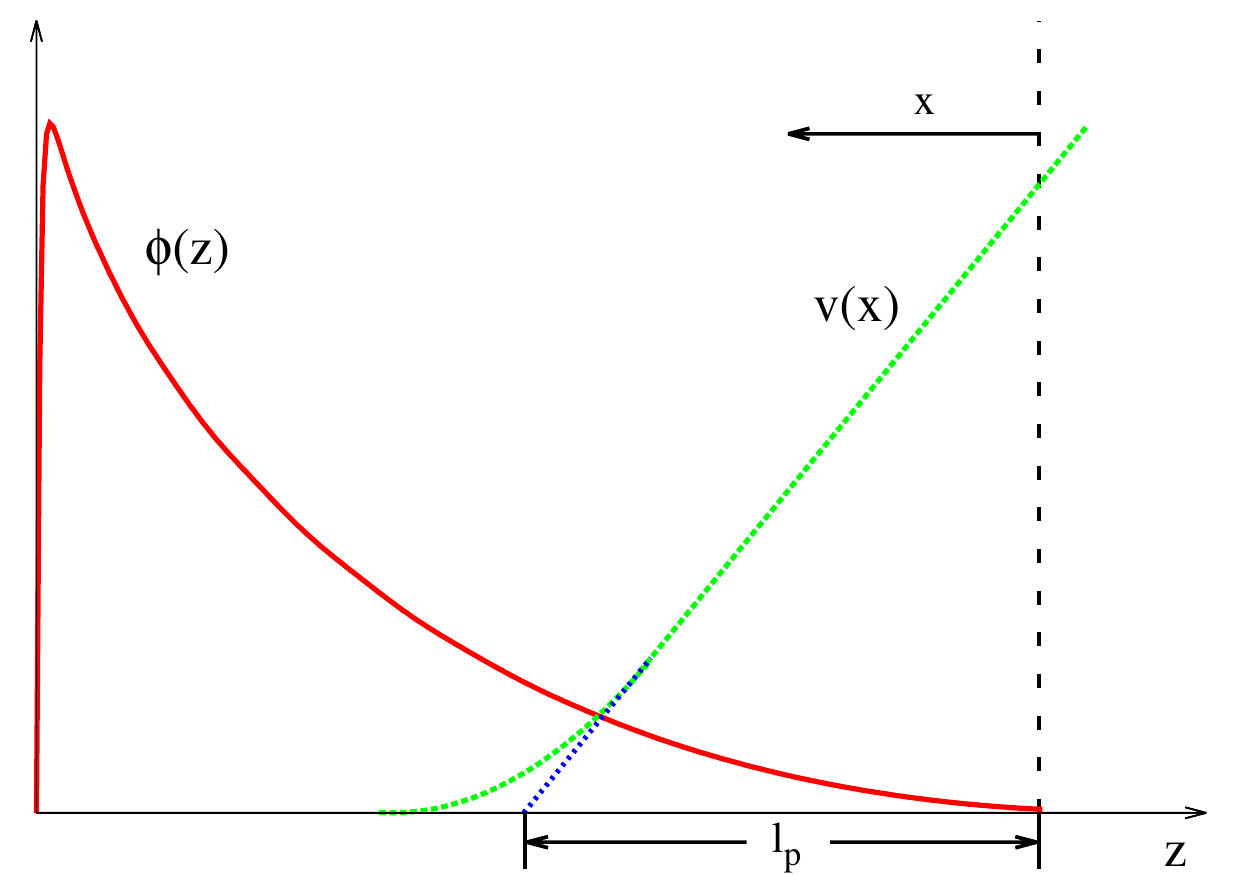}}
  \caption{Sketch of polymer-coated surfaces (a), and construction
  of the hydrodynamic penetration length into one surface (b)}
  \label{fig:hydrodynamic}
\end{figure}

At low shearing rates with $W_{i}\leq0.3$, the brush is almost unperturbed
by the flow \cite{Doyle:1998} and thus certain hydrodynamic parameters
can be linked to equilibrium brush characteristics. We study the problem
of shear flow penetration into an unperturbed brush following Milner
\cite{Milner:hydrodynamic} who evaluated the hydrodynamic penetration
depth for a parabolic monodisperse brush. Our main focus is, of course,
to study the effect of brush polydispersity. 

The notion of the penetration length is essential for understanding the
lubrication forces that oppose the approach of a sphere of radius $R$ towards a
flat surface when they are immersed in a liquid of viscosity $\eta$ (see \fig
\ref{fig9a}).  The force is described by the Reynolds formula 
\begin{equation}
F_{\mathrm{lub}}=-6\pi R^{2}\eta\frac{\dot{D}}{D}
\end{equation}
where $D$ is the closest distance between the surfaces. The Reynolds equation
has been tested for solid surfaces down to surface separations of some 10
molecular diameters \cite{Klein:1996}. When the surfaces are coated with
polymer brushes, the parameter $D$ refers to the separation between the points
where the tangential velocity profile at the brush surface extrapolates to
zero. These points coincides neither with the position of the solid substrate
nor with the outer brush edge, see \fig \ref{fig9a}.

The penetration of a shear flow into a monodisperse brush was calculated by
Milner in Ref.\ \cite{Milner:hydrodynamic} under the assumption that the brush
can be treated as a dilute porous medium, according to the seminal paper by
Brinkman \cite{Brinkman:1947}, and that the inhomogeneous unperturbed brush
density profile introduces a position-varying screening length $\lambda(x)$.
In the absence of pressure gradients, the equation describing the tangential
velocity component $v$ as a function of the distance from the outer edge of
the brush, $x=H-z$, (see \fig \ref{fig9b}) reads: 
\begin{equation}
\frac{d^{2}v}{dx^{2}}=\lambda^{-2}(x)v\label{eq:brinkman}
\end{equation}
Several versions were proposed for the connection between the screening length
and the local monomer density $\phi$. Lai and Binder \cite{Lai:1993} assumed
that the scaling expression for the correlation length in semi-dilute solutions
scales as $\lambda\sim\phi^{-3/4}$, while Milner \cite{Milner:hydrodynamic}
used the correlation length that appears naturally in the SCF picture of a
polymer brush in a good solvent, $\lambda\sim\phi^{-1}$. Suo and Whitmore
\cite{Whitmore:2014} recently suggested yet another form for the relation
$\lambda(\phi)$ based on a free-draining assumption for the brush chains. 
Here we will first adopt Milner's approach in studying the polydispersity 
effects and then extend our results to include the free-draining limit.

In the simplest case of the Alexander brush with a constant density,
the solution decays exponentially: 
\begin{equation}
v(x)=v(0)e^{-x/\lambda},
\end{equation}
and the resulting penetration depth $l_{p}=\lambda=1/\phi_{0}$ 
(where $\phi_{0}$ is the density at the grafting surface) depends 
only on the grafting density and not on the chain length
\cite{Milner:hydrodynamic}, 
\begin{equation}
\label{eq:lp_alexander}
l_{p}\sim\sigma^{-2/3} 
\end{equation}
For a monodisperse brush with a parabolic profile,
the density near the brush edge is approximated by a linear function with the
slope evaluated at $z=H_{\mathrm{mono}}$: 
\begin{equation}
\phi(x)=2x\phi_{0}/H_{\mathrm{mono}}\label{eq:miln_lin_profile}
\end{equation}
The solution is given by \eqn (\ref{eq:solution_miln_mono-1}) in Appendix
\ref{sec:appendix_flow}.  The initial decrease of the flow velocity as it
penetrates the brush is approximately linear, and the hydrodynamic penetration
depth is defined as the depth where this velocity profile extrapolates to zero.
The result for $l_{p}$ is presented below in \eqn (\ref{eq:table_for_length})
(top line). The predicted penetration depth is considerably larger than that
obtained for the Alexander brush model (\eqn (\ref{eq:lp_alexander})), but is
still much smaller than the brush thickness. The same solution is obtained for
the moderately polydisperse brush with a profile close to linear in the whole
range of $z$. The only modification to \eqn (\ref{eq:solution_miln_mono-1}) is
that the slope in the expression for $\phi(x)$ must be replaced according by
$\phi(x)=x\phi_{0}/H=\frac{3}{4}x\phi_{0}/H_{\mathrm{mono}}$, which leads to
the penetration depth given in the middle line of \eqn
(\ref{eq:table_for_length}). In the case of a strongly polydisperse brush with
$\pdi=2$, however, the situation changes. The brush is very tenuous near 
its edge, hence one can expect that the flow penetration may be more 
pronounced. The convex shape of the density profile suggests that a quadratic 
approximation to the brush profile is more appropriate than a linear one. 
Indeed, the exact analytical expression, \eqn (\ref{eq:exp_distr_profile}), 
yields the asymptotic tail density
\begin{equation}\label{eq:tail_strong_poly}
\phi(x)=\phi_{0}\left(\frac{2x}{3\pi H_{\mathrm{mono}}}\right)^{2}
\end{equation}

The solution for the flow profile is given by \eqn (\ref{eq:quad_solution})  
in Appendix \ref{sec:appendix_flow} and yields the hydrodynamic penetration
length shown in the bottom line of the equation below.
\begin{equation}
l_{p}=\left\{ \begin{array}{lll}\label{eq:table_for_length}
1.04\left(\frac{H_{\mathrm{mono}}}{\phi_{0}}\right)^{1/2} \hspace*{-0.5cm}
& \sim N_n^{1/2}\sigma^{-1/6} 
& \textrm{monodisperse }\\
1.7\left(\frac{H_{\mathrm{mono}}}{\phi_{0}}\right)^{1/2} \hspace*{-0.5cm}
& \sim N_{n}^{1/2}\sigma^{-1/6} & \textrm{moderately polydisperse}\\
4.2\left(\frac{H_{\mathrm{mono}}^{2}}{\phi_{0}}\right)^{1/3} \hspace*{-0.5cm}
& \sim N_{n}^{2/3} & \textrm{strongly polydisperse }
\end{array}\right.
\end{equation}

These scaling expressions for the penetration depth can also be derived from
simple qualitative arguments. It is clear that very close to the brush
edge the density is low and the flow can penetrate easily. As the
density increases, the local screening length goes down and the flow
is screened stronger. At some point, the local screening length becomes
comparable to the distance that the flow has penetrated already, 
$x\sim\lambda(x)$.  Beyond that point, the flow is effectively stopped. 
Hence the penetration depth can be estimated from a simple condition: 
\begin{equation}
l_{p}=\lambda\left(\phi\left(l_{p}\right)\right)\label{eq:scaling_for_penetr}
\end{equation}
In the case of the monodisperse or moderately polydisperse brush
with $\phi(x)\sim x\phi_{0}/H_{\mathrm{mono}}$, this leads to 
$l_{p}\sim H_{\mathrm{mono}}/(l_{p}\phi_{0})$ and 
$l_{p}\sim\left(H_{\mathrm{mono}}/\phi_{0}\right)^{1/2}$, which coincides
with \eqn (\ref{eq:table_for_length}) (top and middle lines) 
up to numerical prefactors. Likewise, in the case of the strongly
polydisperse brush, the relation $\Phi(x) \sim x^2 \phi_0/H_{\mathrm{mono}}$
yields $l_p \sim \left(H_{\mathrm{mono}}^2/\phi_0\right)^{1/3}$ in agreement
with \eqn (\ref{eq:table_for_length}) (bottom).

We conclude that generally the hydrodynamic penetration depth $l_{p}$ increases
with increasing polydispersity index. However, for weak to moderate
polydispersities, the increase is relatively small and does not modify the
scaling law $l_{p}\sim N_n^{1/2}\sigma^{-1/6}$.  Strong polydispersity
represented by an exponential chain length distribution with $\pdi=2$ leads to
a considerably larger hydrodynamic penetration depth with a different scaling,
$l_{p}\sim N_n^{2/3}$. 

The above discussion is based on the assumption that the brush can be
treated as a porous medium \cite{Milner:hydrodynamic}, i.e., brush monomers
are taken to act as fixed obstacles on the flow. In reality, monomers
can follow the flow to some extent, and even though this does not change
the density profiles significantly at low shear rates, it does
affect the force balance equation that determines the flow profile in the
brush. In the so-called ``free-draining'' limit where hydrodynamic interactions
are neglected \cite{Doyle:1998}, the drag force due to flow-brush interactions
becomes linear in the monomer density. A modified Brinkmann equation that 
conforms to the free-draining model was recently introduced in 
Ref.\ \cite{Whitmore:2014}:
\begin{equation}
\frac{d^{2}v}{dx^{2}}=\frac{\zeta}{\eta}\frac{\phi(x)}{1-\phi(x)}v(x)=\lambda^{-2}(x)\, v
\end{equation}
Here the screening length is now related to the monomer density as
$\lambda\sim\left(\frac{1-\phi}{\phi}\right)^{1/2}$. Considering only the low
monomer density limit $\phi(x)\ll1$ and omitting prefactors of order 1, we
obtain
\begin{equation}
\frac{d^{2}v}{dx^{2}}=\phi(x)\ v\label{eq:free draining}.
\end{equation}
Repeating the calculations for the four cases discussed above yields
the following expressions for the penetration depth into free-draining
brushes: 
\begin{equation}
l_{p}=\left\{ \begin{array}{lll}
\left(\frac{H_{\mathrm{mono}}}{\phi_{0}}\right)^{1/3} & \sim N_n^{1/3}\sigma^{-1/9} & \textrm{monodisperse }\\
\left(\frac{H_{\mathrm{mono}}}{\phi_{0}}\right)^{1/3} & \sim N_{n}^{1/3}\sigma^{-1/9} & \textrm{moderately polydisperse}\\
\left(\frac{H_{\mathrm{mono}}^{2}}{\phi_{0}}\right)^{1/4} & \sim N_{n}^{1/2} & \textrm{strongly polydisperse }
\end{array}\right.
\end{equation}

All these scaling results still follow from \eqn (\ref{eq:scaling_for_penetr}),
using $\lambda(\phi)=\phi^{-1/2}$. Due to the increased
drag force on the solvent in the free-draining case, it is more difficult
for the flow to penetrate the brush compared to the case of partially screened
hydrodynamics.

We should stress again that our treatment, which was based on the assumption
that the brush profile is not perturbed by the shear flow, can only be used for
small shear rates. This restriction is particularly relevant in the case of
strongly polydisperse brushes which are very tenuous and thus susceptible to
small external forces near the outer edge.  Whether the treatment presented
above is valid in under prevalent experimental conditions has been a subject of
some debate \cite{Klein:1996}.  

The behavior of brushes in shear flow has been studied extensively by various
versions of MD, dynamic MC, and DPD simulations.  Some MD simulations
\cite{Doyle:1998} have employed the free-draining approximation to evaluate the
drag forces applied to brush chains. In contrast, Wijmans and Smit
\cite{Wijmans:2002} carried out DPD simulations with explicit solvent that did
not make any assumptions regarding the velocity profile in the polymer layer.
Their results can be used to check the validity of theories that describe the
solvent flow, such as the Brinkman equation and, in particular, the dependence
of the position-dependent screening length on the local monomer density.
According to these simulations for rather short brush chains with $N=20$ and
several grafting densities, the data on the screening length $\lambda$ as a
function of the monomer density collapse onto a master curve, giving
$\lambda\sim\phi^{-0.55}$.  This is in better agreement with the free-draining
treatment than with the non-draining relation in the absence of swelling
effects $\lambda\sim\phi^{-1}$ utilized by Milner \cite{Milner:hydrodynamic}.

A direct DPD simulation of the flow penetration into a brush-coated surface and
the penetration depth was recently carried out by Deng et.al. \cite{Deng:2012}.
The brush was monodisperse. With the flow velocity gradients used in the
simulations, the brush density profile was almost unperturbed, which
corresponded to Weissenberg numbers of less than $W_i=1.$ The observed
penetration depth was considerably smaller than the total brush height (roughly
1/10) but not small compared to atomic sizes (about 5 nm after conversion of
the model parameters to ordinary units). It was shown that the penetration
depth decreases with increasing grafting density, $\sigma$, although the change
is not large. The scaling exponent for the $l_{p}(\sigma)$ dependence was not
estimated, and the length of the brush chains was not varied.

We close this section with a brief discussion on the total shear stress
experienced by the brush-coated surface. The average drag force exerted by the
flowing solvent on the brush-coated surface depends on the velocity profile,
which, in turn, is determined by details of the brush-flow interaction.  Under
stationary conditions, the force balance requirement results in a simple
identity that could provide a useful test to check the consistency of the
numerical work. Imagine a pair of opposing surfaces in Couette geometry, one of
which is brush-coated and the other is bare. The shear stress acting on the
bare surface is completely determined by the velocity gradient at the boundary
$\eta\dot{\gamma}$. This gradient is the same across the gap where only pure
solvent is flowing, down to the outer brush edge and can be accurately measured
in simulations. The total drag per unit area of the brush-coated surface must
always be given by the same expression, irrespective of the brush parameters or
the detailed way of treating hydrodynamics. Thus the value of the stationary
velocity gradient outside the brush encodes all the relevant information on the
brush-flow interaction.

\section{Summary: Main features of polydisperse brushes}\label{sec:summary}

The main focus of the present paper was to systematically compare the
equilibrium and hydrodynamic properties of monodisperse and polydisperse
brushes.  We first recapitulate some very general results that have first been
obtained by Milner, Witten, and Cates \cite{Milner_Polydisp} for arbitrary
molecular mass distributions, but have remain underappreciated in the current
literature in our opinion:

1.) Under good solvent conditions and in the strong crowding limit,
the monomer density at the grafting surface, $\phi_{0}$, is determined
only by $\sigma$ and the excluded volume parameter, $v$: 
\[
\phi_{0}=\frac{3}{2}\left(\frac{\pi\sigma v}{2}\right)^{2/3}
\]
and depends neither on the average chain length, nor on the shape of the chain
length distribution. In simulations, one observes some decrease in the brush
density near the surface for larger polydispersities which is a finite-$N$
effect that would disappear in the asymptotic limit.

2.) Under the same conditions, the brush height always scales as: 
\[
H\sim N_{n}\sigma^{1/3}
\]
irrespective of the particular shape of the chain length distribution and of
the polydispersity index. The prefactor, however, is a function of the chain
length distribution and typically increases with the polydispersity parameter.

Turning to our own results, in the following, we summarize the changes in the
brush properties produced by polydispersity effect. In the present paper we
have mostly focussed on the Schultz-Zimm family of molecular mass
distributions, and hence a polydisperse brush is uniquely characterized by 3
parameters: the number-averaged molecular mass (or polymerization index),
$N_{n}$, the grafting density, $\sigma$, and the polydispersity index $\pdi$ .

3.) At fixed $N_{n}$ and $\sigma$, the brush height increases with the
polydispersity parameter, which can be approximately described by a very simple
expression as demonstrated in \fig \ref{fig:brush_heights}. 

4.) With increasing polydispersity, the shape of the brush
density profile systematically changes from the concave down parabolic
shape to a convex shape. At moderate polydispersity, $\pdi\backsimeq1.15$
the profile is close to linear with a constant slope 
(see \fig \ref{fig:profiles}).

5.) A monodisperse brush is characterized by anomalous chain end fluctuations
$\mathrm{var}Z_{N}\sim H^{2}$ due to the fact that the end monomer of each
chain explores all the space within the brush thickness. In a polydisperse
brush, the end of each chain fluctuates around its own mean position. With 
increasing polydispersity, the anomalous fluctuations are suppressed as
illustrated by \fig \ref{fig:The-maximal-fluctuations}, so that eventually the
fluctuations of a typical chain recover the ideal behavior
$\mathrm{var}Z_{N_{n}}\sim N_{n}$ .

At least two effects are expected as consequences of this change. The first is
related to chain relaxation dynamics. A monodisperse unentangled brush exhibits
very large relaxation times $\tau\sim N_n^{3}$; however, we expect Rouse
relaxation times $\tau\sim N_{n}^{2}$ to be restored even in moderately
polydisperse brushes. \RE{For longer chain lengths $N_n$, entanglement effects
become important. A recent study on dense monodisperse polymer brushes showed
that the relaxation time scales exponentially with the chain length $N_n$ with
the prefactor $N_n^3$, i.e., $\tau\propto N_n^3\exp(N_n/N_e)$ where $N_e$ is the
entanglement degree \cite{entanglement_brush}. We would expect that when the 
entanglement is involved, the relaxation time might be reduced to 
$\tau\sim N_n^2\exp(N_n/N_e)$ even in moderately polydisperse brushes.}
The second consequence relates to the effect reported in
Refs.\ \cite{Merlitz:2008, Romeis:2012} as ``surface instability'' of a
monodisperse brush: The modification of the end-group of some brush chains so
that it becomes sensitive to small variations in the solvent quality results in
a possibility of an abrupt transition for the modified chains either ``hiding''
inside the brush or exposing the end-group at the outer brush edge. The
transition is clearly linked to the anomalously high susceptibility of chain
ends in a monodisperse brush. As the latter is strongly suppressed by
polydispersity, we expect the surface instability \RE{to be affected as well}.

6.) Hydrodynamic flow near a brush-coated surface penetrates into the
brush. The penetration depth, $l_{p}$, increases with the polydispersity
since the brush becomes more tenuous at the edge. For moderate polydispersities
the effect is not very large, but for $\pdi=2$ we
predict a stronger scaling $l_{p}\sim N_{n}^{2/3}$ as opposed to
$l_{p}\sim N_n^{1/2}\sigma^{-1/6}$ in the monodisperse case.

We hope that the relatively simple picture of polydisperse brushes, their
internal structure and related properties, which we have presented here, may be
useful for researchers developing new methods of brush synthesis and
applications of brush-coated surfaces.


\bigskip
\begin{center}
\textbf{ACKNOWLEDGMENTS}
\end{center}

Financial support by the Deutsche Forschungsgemeinschaft (Grants No. SCHM
985/13-2) is gratefully acknowledged. S. Qi acknowledges support from the
German Science Foundation (DFG) within project C1 in SFB TRR 146. Simulations
have been carried out on the computer cluster Mogon at JGU Mainz.

\appendix
\section{Numerical SCF theory}\label{SCF_1d_appendix}

In the SCF approach \cite{SCF_R1,SCF_book}, the configurational partition
function is converted to path integrals over the fluctuating fields by invoking
functional delta constraints. One obtains $\mc Z=\int\mc
D\omega\mc D\rho_te^{-\beta\mc F}$ with the free energy functional expressed as
\begin{eqnarray}
\beta\mc F&=&\frac{v}{2}\int d\mb r\phi^2-\int d\mb r\omega\phi-\sum_{\alpha=1}^{n_b}\ln Q_\alpha[\omega]
\end{eqnarray}
where $\phi$ is the total fluctuating density, $\omega$ is the corresponding
conjugate auxiliary potential, and $Q_\alpha$ is the single chain partition
functions for the $\alpha$-th chain. For a system with macroscopic volume $V$,
we assume that the sum over all brush chains can be replaced by a continuous
integral weighted by a proper continuous chain length distributions, which
means
\begin{equation}
\sum_{\alpha=1}^{n_b}\ln Q_\alpha[\omega]\simeq n_b\int_0^\infty dNP(N)\ln Q(N,\omega)
\end{equation}
where $Q$ is the partition function of a brush chain with chain length $N$, and
$P(N)$ is defined as the quenched chain length distribution function normalized
to unity $\int_0^\infty dNP(N)=1$. Extremizing the free energy functional with
respect to the fluctuating fields, i.e., the density $\phi$ and the potential
$\omega$, we obtain a set of closed SCF equations
\begin{eqnarray}
\omega&=&v\phi\\
\phi&=&n_b\int_0^\infty dNP(N)\frac{1}{Q(N)}\int_0^Ndsq(\mb r,s)q^\dagger(\mb r,N-s)
\nonumber
\end{eqnarray}
where the single chain partition function can be calculated from the
propagators, i.e., $Q_N=\int d\mb rq^\dagger(\mb r,N)$. The propagators
$q^\dagger$ and $q$ satisfy the same modified diffusion equation
\begin{equation}
\frac{\partial q(\mb r,s)}{\partial s}=\frac{b^2}{6}\nabla^2q(\mb r,s)-\omega(\mb r)q(\mb r,s)
\end{equation}
but with different initial conditions. We use the initial condition $q(\mb
r,0)=1$, and $q_\dagger(\mb r,0)=\delta(z)$, i.e., one end of each chain is
free, while the other end is grafted at $z=0$ with mobile grafting point. Since
the grafting substrate is impenetrable for all chains, we implement Dirichlet
boundary condition there.

Inserting the SZ distribution function $P(N)$ into the density $\phi$, we obtain
\begin{eqnarray}
\phi&=&\frac{n}{\Gamma(k)}\int_0^\infty dxx^{k-1}e^{-x}\nonumber\\
&\times&\frac{1}{Q[xN_n/k]}\int_0^{xN_n/k}dsq(\mb r,s)q^\dagger(\mb r,xN_n/k-s)
\end{eqnarray}where $x=kN/N_n$. In practice, the above infinite integration is approximated by the Gauss-Laguerre quadrature \cite{numerical_recipes}, which has the form of
\begin{equation}
\int_0^\infty dxx^{k-1}e^{-x}f(x)=\sum_{j=1}^{n_G}w_jf(x_j)
\end{equation}
where the abscissae ($x_j$) and weights ($w_j$) can be evaluated numerically.
This quadrature normal converges very rapidly. For example, for a polymer
blend, $n_G\lesssim 10$ is enough to generate converged interfaces. For the
present brush system, we found that several ten points are necessary to obtain
smooth and accurate brush density profiles. With a smaller number of sampled
chain lengths, the obtained brush densities are not so smooth. In the present
paper, we adopt $n_G=60$.  The modified diffusion equation is solved in real
space using the Crank-Nicolson scheme, which is a finite difference method. The
SCF equations are solved iteratively by updating the potential $\omega$ with a
simple mixing scheme. In the main text we consider only the one dimensional 
system. 

\section{Structure of polydisperse brushes: theory}\label{sec:theory-appendix}

\subsection{Analytical SCF theory in the strong stretching approximation}
\label{sec:theory_MWC}

An important quantity that needs to be specified in order to characterize
the structure properties of a polydisperse polymer brush is the self-consistent
potential $\omega(z)$. At the mean field level, in good solvent, this potential is
proportional to the monomer density with the proportionality
constant being the excluded volume parameter.

We begin with briefly describing the mean-field approach we use to
evaluate the potential. It is based on the MWC theory \cite{Milner_Polydisp}.
Our goal is to derive an expression for the potential as a function
of the distance from the substrate, i.e., the spatial coordinate $z$.
We assume that the potential is a monotonically decreasing function
of $z$ with its maximum a $z=0$, where the monomer density is high,
and that it vanishes at the brush edge $z=H$. The potential can then
be written as $\omega=A-U$, where $U$ is a monotonically increasing
function with $U=0$ at $z=0$ and $U=A$ at $z=H$. For $0<z<H$
there hence exists a unique relation between $z$ and $U$.

Apart from the mean-field approximation, we also adopt the strong
stretching approximation, i.e., we assume that chains never loop back,
and take the asymptotically long chain limit whereby fluctuations
can be neglected. As demonstrated in the MWC theory \cite{Milner_Polydisp},
the density and the potential are connected through the free end distribution
$\Phi_{e}(z)$ via 
\begin{equation}
\phi(z)=\Big(\frac{3}{2}\Big)^{1/2}\int_{U(z)}^{A}dU'\frac{\epsilon(U')}{\sqrt{U'-U(z)}}
\end{equation}
in this limit, where we have defined $\epsilon(U)=\Phi_{e}(z)\: dz/dU$.
Since we have $\phi(z)=A-U(z)$ for a brush at the mean field level
(note again that we set $v=1$), this equation implies the relation
$\epsilon(U)=\sqrt{\frac{8}{3\pi^{2}}}(A-U)^{1/2}$. From $\epsilon(U)$,
we can define a cumulative grafting density $\sigma_{c}(z)$, corresponding
to the number of polymer chains per unit area with chain end position
at distance smaller than $z(U)$ from the substrate, i.e., 
$\sigma_c[U]=\int_{0}^{U}dU'\epsilon(U')$.
Inserting the expression for $\epsilon$, we can express the cumulative
grafting density as a function of $U$, 
\begin{equation}\label{eq:sigma_U}
\sigma_{c}(U)=\sqrt{\frac{8}{3\pi^{2}}}\frac{2}{3}A^{3/2}\Big[1-\Big(1-\frac{U}{A}\Big)^{3/2}\Big].
\end{equation}

As the total grafting density for all chains is fixed to be $\sigma$,
which requires $\sigma_{c}(A)=\sigma$, we have 
$\sigma=\sqrt{\frac{8}{3}}\frac{2}{3\pi}A^{3/2}$,
or in other words,
\begin{equation}\label{eq:A_sigma}
A=\frac{3}{2}\big(\frac{\pi\sigma}{2}\big)^{2/3}.
\end{equation}
This implies that the monomer density at the grafting surface $z=0$
is only determined by the total grafting density, independent of the
polydispersity.

In a polydisperse brush, according to the strong segregation approximation,
chains with different lengths are completely segregated, and
there is a unique relation between the chain length $N$, the
position of the end monomer $z$, and the corresponding value of $U(z)$.
Hence we can associate $\sigma_{c}(U)$ with the cumulative grafting
density $\sigma_{c}(N)$ of chains with length shorter than $N$,
and from \eqn (\ref{eq:sigma_U}), we can express $U$ as a function
of $\sigma_{c}(N)$, i.e.,
\begin{equation}\label{eq:UN}
U(N)=A\Big[1-\Big(1-\frac{\sigma_c(N)}{\sigma}\Big)^{2/3}\Big]
\end{equation}

On the other hand, the inverse function, $N(U)$, obeys 
\begin{equation}\label{eq:NU}
N(U)=\int_{0}^{U}dU'\frac{dN}{dU'}=\sqrt{\frac{3}{2}}\int_{0}^{U}dU'\frac{dz}{dU'}\frac{1}{\sqrt{U-U'}}
\end{equation}
in the strong stretching theory \cite{Milner_Polydisp}, and one can
easily see that this equation is fulfilled if 
\begin{equation}\label{eq:zU}
\frac{dz}{dU}=\sqrt{\frac{2}{3\pi^{2}}}\int_{0}^{U}dU'\frac{dN}{dU'}\frac{1}{\sqrt{U-U'}}.
\end{equation}From these equations we can obtain $U(z)$ is $N(U)$ is explicitly
known.

Finally, \eqn (\ref{eq:zU}) may be integrated to find a general
expression for the equilibrium height:
\begin{equation}
H=\frac{2}{\pi}\sqrt{\frac{2}{3}}\int_{0}^{N_\mathrm{max}}dn\left(A-U(n)\right)^{1/2}
\end{equation}
Since every chain length distribution with a finite first moment
can be written in the form $P(N)=N_{n}^{-1}\psi\left(\frac{N}{N_{n}}\right)$,
it follows from \eqn (\ref{eq:UN}) that $U(n)/A$ is a function of the
reduced variable $\frac{n}{N_{n}}$ only, the values of the function
being in the range [0,1]. Rescaling the integration variable, $n=N_{n}t$ we arrive at
\begin{equation}
H=N_{n}\frac{2}{\pi}\sqrt{\frac{2A}{3}}\int_{0}^{\frac{N_{max}}{N_{n}}}dt\left(1-\frac{U(t)}{A}\right)^{1/2}
\end{equation}
Substituting $A$ from \eqn (\ref{eq:A_sigma}) we obtain 
\begin{equation}\label{eq:H_N_sigma}
H=\left(\frac{2}{\pi}\right)^{2/3}N_{n}\sigma^{1/3}\int_{0}^{\frac{N_\mathrm{max}}{N_{n}}}dt\left(1-\frac{U(t)}{A}\right)^{1/2}
\end{equation}The integral is a pure number that is determined solely by the shape
of the chain length distribution but does not depend on $\sigma$ or
$N_{n}$. Thus the scaling relation 
\begin{equation}
H\sim N_{n}\sigma^{1/3}
\end{equation}is generally valid for any non-pathological chain length distribution.
For a monodisperse brush the integral is reduced to $\int_{0}^{1}dt\left(1-\theta(t-1)\right)^{1/2}=1.$
For any other chain length distribution it will have the meaning of
$\frac{H}{H_{\mathrm{mono}}}$. As another simple illustration, we
note that for an exponential distribution density, \eqn (\ref{eq:UN_exp})
leads to the explicit form $\int_{0}^{\infty}dt\exp\left(-\frac{t}{3}\right)=3$.

\subsection{Monodisperse brush with a vanishing fraction of minority chains of different length}
\label{sec:theory_mono}

The monodisperse brush corresponds to the very special limit where
$N(U)$ in \eqn (\ref{eq:NU}) is a constant, $N(U)\equiv N_{n}$.
Even though the assumption of chain length segregation obviously breaks
down in this limit, and the function $N(U)$ cannot be inverted, all
equations of the previous section except \eqn (\ref{eq:UN}) still hold
(with $dN/dU=N_{n}\:\lim_{\epsilon \to 0^+}\delta(U-\epsilon)$) and 
\eqn (\ref{eq:zU}) immediately
gives $U(z)=-\frac{3\pi^{2}}{8N_{n}^{2}}z^{2}$.

Here we summarize the main structural properties of monodisperse brushes.
The density profile is given by: 
\begin{equation}\label{eq:mono_profile}
\phi(z)=\phi(0)\left(1-\frac{z^{2}}{H_{\mathrm{mono}}^{2}}\right)
\end{equation}
where $\phi(0)=\frac{3}{2}\left(\frac{\pi\sigma}{2}\right)^{2/3}$.
The brush height 
\begin{equation}\label{eq:H_mono}
H_{\mathrm{mono}}=\left(\frac{4v\sigma}{\pi^{2}}\right)^{1/3}N_{n}
\end{equation}
scales linearly with the chain length $N_{n}$. The brush density
generates the chemical potential profile: 
\begin{equation}\label{eq:brush_potential}
\omega(z)=\frac{3\pi^{2}}{8N_{n}^{2}}(H_{\mathrm{mono}}^{2}-z^{2})
\end{equation}
Next we introduce a vanishing fraction of minority chains of
different length, $N$, into the otherwise monodisperse brush. The
Green's function of a minority chain is given by the known solution
of the Edwards' equation 
\begin{equation}\label{eq:Edwars}
\frac{\partial G(z,s)}{\partial s}=\frac{1}{6}\frac{\partial^{2}G(z,s)}{\partial^{2}z}-\omega(z)G(z,s)
\end{equation}
in the presence of a parabolic potential \eqn (\ref{eq:brush_potential})
and an impenetrable inert grafting surface \cite{Skvortsov:1997}:
\begin{equation}\label{eq:Green_parabolic}
G(z,N)=B\: z\exp\left[-\frac{3\pi}{4N_{n}}\cot\left(\frac{\pi N}{2N_{n}}\right)z^{2}\right]
\end{equation}
Here $B$ is a normalization factor. If minority chains are longer
or equal to the mean length of the majority chain, $N\geq N_{n}$,
the Green's function \eqn (\ref{eq:Green_parabolic}) shows an unbounded
monotonic increase with $z$ since the solution refers to a potential
extending to infinity and an ideal Gaussian chain with unlimited extensibility.
Even chains slightly shorter than $N_{n}$ are predicted to extend
beyond the brush edge at $z=H_{\mathrm{mono}}$. However, the brush
edge modifies the potential at $z>H$ dramatically which roughly amounts
to a truncation of the Green's function \eqn (\ref{eq:Green_parabolic})
at $z=H_{\mathrm{mono}}$ for chains with $N\leq N_{n}$. Comparison
with numerical SCF calculations shows that for longer chains, $N\geq N_{n}$,
the truncated Green's function describes well the position of the
$N_{n}$-th monomer, while the last $N-N_{n}$ monomers have a mushroom-like
configuration with the average end-to-end distance of $[\pi(N-N_{n})/6]^{1/2}$.
Obtaining the full expression for the average distance of the free
end $Z_{e}(N)$ is rather cumbersome, but most of the curve can be
described by two simple limits:

\begin{equation}
Z_{\mathrm{e}}(N)=\left\{ \begin{array}{ll}
\left[\frac{N_{n}}{3}\tan\left(\frac{\pi N}{2N_{n}}\right)\right]^{1/2}, 
  & N<N_{n}\\
H_{\mathrm{mono}}-\frac{4N_{n}^{2}}{3\pi^{2}H(N-N_{n})}
  +\left[\frac{\pi(N-N_{n})}{6}\right]^{1/2}, & N>N_{n}
\end{array}\right.\label{eq:z_minority}
\end{equation}
where the difference $|N-N_{n}|$ must be at least a few units for
the approximation to be meaningful. Similarly, the variance of the
end-monomer position var$Z_{N}$ is evaluated using the truncated
Green's function for $N\leq N_{n}$, while for $N>N_{n}$ an extra
term $\frac{(4-\pi)}{6}(N-N_{n})$ accounting for the terminal subchain
with a mushroom configuration is added. For the fluctuations of the
end-monomer positions we have two simple limits: 
\begin{equation}
\mathrm{var}Z_{N}=\left\{ \begin{array}{ll}
\frac{(4-\pi)}{3\pi}N_{n}\tan\Big(\frac{\pi N}{2N_{n}}\Big), & N<N_{n}\\
\frac{16N_{n}^{4}}{9\pi^{4}H_{\mathrm{mono}}^{2}(N-N_{n})^{2}}+\frac{(4-\pi)}{6}(N-N_{n}), & N>N_{n}
\end{array}\right.\label{eq:variation_minority_chains}
\end{equation}

\subsection{Weakly polydisperse brush}\label{sec:theory_field_weakly_p}

A monodisperse brush is characterized by the relation $\frac{U(N)}{A}=\Theta(N-N_{n})$.
To introduce an analytically tractable model of weakly polydisperse
brushes we replace the $\Theta$-function by a function which is reminiscent
of the well-known low-temperature limit of the Fermi-Dirac hole occupancy:
\begin{equation}
\label{eq:un}
\frac{U(N)}{A}=1-\frac{1}{\exp\left(\frac{1}{\Delta}\left(\frac{N}{N_{n}}-1\right)\right)+1}
\end{equation}where $\Delta$ is a small parameter analogous to the reduced temperature
$\frac{T}{T_{F}}$ in the Fermi gas: It gives the relative width of
the ``smoothing''. In the language of the chain length distribution
density, $P\left(\frac{N}{N_{n}}\right)$, $\Delta$ also defines
its width. For a narrow distribution of relative width $\Delta,$
$N_{w}\sim N_{n}\left(1+\Delta^{2}\right)$, leading to 
\begin{equation}
\Delta\sim\left(\pdi-1\right)^{1/2}
\end{equation}
Solving \eqn (\ref{eq:un}) for $N(U)$ and substituting into \eqn (\ref{eq:zU})
gives:
\begin{eqnarray}\label{eq:dzdu fermi}
\frac{dz}{dU}&=& \sqrt{\frac{2}{3A}} \frac{N_n}{\pi} \frac{1}{\sqrt{U}}
  \Bigg[ 1 + \Delta \Big(\log\frac{4U}{A}\nonumber\\
&& + \;  2\sqrt{\frac{U}{A-U}}\arctan \big(\frac{U}{A-U}\big) \Big)\Bigg]
\end{eqnarray}
(note that in this case, the lower bound in the integral (\ref{eq:zU}) and
$\int dU'\frac{dz}{dU'}$ cannot be set to zero, but must be chosen
$U(0)\approx \exp(-\frac{1}{\Delta}) \neq 0$).

Our main focus is to estimate the second derivative $\frac{d^{2}U}{dz^{2}}$
which is expressed as
\begin{equation}
\frac{d^{2}U}{dz^{2}}=-\left(\frac{dz}{dU}\right)^{-3}\frac{d}{dU}\left(\frac{dz}{dU}\right)
\end{equation}up to terms linear in the small parameter 
\begin{equation}
\frac{d^{2}U}{dz^{2}}=3\left(\frac{\pi}{2N_{n}}\right)^{2}\left(1-\Delta\psi\left(\frac{U}{A}\right)\right),
\end{equation}where $\psi\left(\frac{U}{A}\right)$ is obtained from 
\eqn (\ref{eq:dzdu fermi}) and positive for $U/A > 0.07$. 
Hence the curvature of
the mean-force potential is indeed reduced, the relative reduction
being linear in $\Delta$. Evaluating the function $\psi$ close to
the middle point of the brush, $U/A=1/2$ we obtain an approximation
for the typical curvature of the potential:
\begin{equation}
\frac{d^{2}U}{dz^{2}}=3\left(\frac{\pi}{2N_{n}}\right)^{2}\left(1-c\Delta\right)
\end{equation}
with $c \approx 6$. 
Now the fluctuations of a chain with $N=N_{n}$ can be obtained from
the Green's function \eqn (\ref{eq:Green_parabolic}) modified to account
for the reduced coefficient in the parabolic potential:
\begin{equation}
G(z,N_{n})=B\: z\exp\left[-\frac{3\pi}{4N_{n}}\cot\left(\frac{\pi}{2}\left(1-\frac{c}{2}\Delta\right)\right)z^{2}\right]
\end{equation}In the limit of $\Delta\ll1$ the fluctuations scale as
\begin{equation}
\mathrm{var}Z_{N_{n}}\sim N_{n}\Delta^{-1}\sim N_{n}\left(\pdi-1\right)^{-1/2},
\end{equation} as long as they don't exceed those in a purely monodisperse brush.

\subsection{Moderately polydisperse brush with exactly linear density profile}
\label{sec:theory_linear}

The density profile of a brush formed by chains with SZ distribution
at $\pdi=1.143$ ($k=7$), is reasonably close to a
linear shape. From a theoretical point of view this presents a considerable
advantage since the Green's function of an ideal chain placed in a
linear potential is known \cite{Mansfield}. Equally fortunately,
one can also find a closed-form analytical description of a polydisperse
brush with exactly linear profile within the SCF strong-stretching
approximation. We write the linear density profile as 
\begin{equation}
\phi_{\mathrm{lin}}(z)=\phi_{0}-fz\label{eq:linear_profile}
\end{equation}
where we have discussed that $\phi_{0}\equiv A$ has nothing to do
with the polydispersity. From the density profile, the height of the
brush can be defined as $H=\phi_{0}/f$. The slope $f$ can be specified
according to the fact that the total number of monomers in the system
are fixed to be $\sigma N_{n}$, i.e., one can get $f=\phi_{0}^{2}/(2\sigma N_{n})$.
According to our previous notation, we have $U=fz$, and thus $dz/dU=1/f$.
According to \eqn (\ref{eq:NU}), we obtain $N(U)=\sqrt{6U}/f$, from
which we have $U=f^{2}N^{2}/6$. Subsequently, we can evaluate the
cumulative grafting density according to \eqn (\ref{eq:sigma_U})
\begin{equation}
\frac{\sigma_c(N)}{\sigma}=\left[1-\left(1-\frac{f^{2}N^{2}}{6A}\right)^{3/2}\right]
\end{equation}
The derivative of $\sigma_c(N)/\sigma$ with respect to $N$ gives the
chain length distribution function 
\begin{equation}
P_{\mathrm{lin}}(N)=\frac{3N}{N_{\mathrm{max}}^{2}}\left[1-\frac{N^{2}}{N_{\mathrm{max}}^{2}}\right]^{1/2}\label{eq:CLD_linear_profile}
\end{equation}
where the cut-off chain length $N_{\mathrm{max}}$ is defined as $N_{\mathrm{max}}=\sqrt{6A}/f=\frac{16}{3\pi}N_{n}$.
It can be checked that this chain length distribution is normalized
to 1, and the number-averaged chain length is equal to $N_{n}$ as
expected. The polydispersity index can be directly calculated as $\pdi=1.153$,
which is quite close to that of the SZ distribution with $k=7$ although
the actual shape differs as demonstrated in 
\fig \ref{fig:moderate_polydisperse}(a)
and \fig \ref{fig:moderate_polydisperse}(d) (insets). In terms of
$H_{\mathrm{mono}}$ with the same parameter $\sigma$ and $N_{n}$ given
by \eqn (\ref{eq:H_mono}), the brush thickness of the moderate polydisperse
brush can be expressed as $H=\frac{4}{3}H_{\mathrm{mono}}$. It is
a puzzling coincidence that the form of the chain length distribution
given by \eqn (\ref{eq:CLD_linear_profile}) is the same as the end
monomer density distribution in a monodisperse brush.

A linear density profile generates a uniform constant stretching force
acting on each monomer, very similar to a gravitational field. The
Green's function of a chain placed in a uniform force field of strength
$f$ is a solution of the Edwards' equation (\eqn (\ref{eq:Edwars})) with
$\omega(z)=-fz$, which was obtained in Ref.\ \cite{Mansfield}: 
\begin{equation}\label{eq:Green_function_at_k7-1}
G(z,N)=\left(\frac{2\pi N}{3}\right)^{-1/2}\exp\left[\frac{f^{2}N^{3}}{18}-\frac{3}{2N}\left(z-\frac{f\, N^{2}}{6}\right)^{2}\right]
\end{equation}
A closely related solution in the context of quantum mechanics (propagator
for a non-relativistic particle in an accelerator) has also been known
\cite{Feynman}. The polymer Green's function was used to describe
properties of a chain in a viscous flow or of a charged chain in a
uniform electric field. It is quite striking that a polydisperse brush
turns out to be another possible source of a linear potential. The force $f$
is to be identified with the slope of the linear density profile:
\begin{equation}
f=\frac{9\pi}{16N_{n}}\left(\frac{\pi\sigma}{2}\right)^{1/3}\label{eq:force_in_linear_field}
\end{equation}
For a semi-quantitative comparison of the theory with the MC data,
the effect of the impenetrable grafting surface and the cut-off of
the field for $z>H$, i.e., beyond the brush edge, must be taken into
account. This is done in manner similar to the monodisperse case:
\begin{equation}
G(z,N)=B'\: z\ \exp\left[-\frac{3}{2N}\left(z-\frac{fN^{2}}{6}\right)^{2}\right],\qquad z\leq H,\label{eq:cut_off_effect}
\end{equation}
where $B'$ is a normalization constant

Like in the monodisperse case, the Green's function (\eqn (\ref{eq:Green_function_at_k7-1}))
is used to evaluate $Z_{e}(N)$ and var$Z_{N}$ for chains with $N\leq N_{\mathrm{max}}$.
For longer chains, these values are interpreted as referring to the
$N_{\mathrm{max}}$-th monomer, and the contributions due to the mushroom-like
tail of length $(N-N_{\mathrm{max}})$ are added in the same way as
in Sec.\ \ref{sec:theory_mono}, except that $N_{n}$ is replaced
by $N_{\mathrm{max}}$ here.

\subsection{Strongly polydisperse brush}\label{sec:theory_exp}

The SZ distribution with $\pdi=2$ ($k=1$ ) has a
simple exponential shape: 
\begin{equation}
P(N)=\frac{1}{N_{n}}\exp\left(-\frac{N}{N_{n}}\right)\ \label{eq:ZimmShulz_at_k1}
\end{equation}
This allows a closed-form analytical solution within the framework
of the theory developed by Milner, Witten, and Cates (MWC) \cite{MWC}.
The cumulative grafting density (distribution) $\sigma_{c}(N)$ which
plays a central role in the theory can be calculated directly as 
\begin{equation}
\frac{\sigma_{c}(N)}{\sigma}=\int_{0}^{N}dN'P(N')=\left[1-\exp\left(-\frac{N}{N_{n}}\right)\right]
\end{equation}
The chain length as a function of $U$ can be obtained according to
\eqn (\ref{eq:UN})
\begin{equation}
U(N)=A\left[1-\exp\left(-\frac{2N}{3N_{n}}\right)\right],\label{eq:UN_exp}
\end{equation}
from which we have the derivative 
\begin{equation}
\frac{dN}{dU}=\frac{3}{2}\frac{N_{n}}{\left(A-U\right)}.
\end{equation}
Inserting this above equation into \eqn (\ref{eq:zU}) and performing
the integration, we get 
\begin{equation}
\frac{dz}{dU}=\sqrt{\frac{6N_{n}^{2}}{\pi^{2}}}(A-U)^{-1/2}\arcsin\sqrt{\frac{U}{A}}.
\end{equation}
Integrating this equation, we get 
\begin{equation}
z(U)=\frac{6N_{n}}{\pi}\Big(\frac{2A}{3}\Big)^{1/2}\left[\sqrt{\frac{U}{A}}-\sqrt{1-\frac{U}{A}}\arcsin\sqrt{\frac{U}{A}}\right].\label{eq:zU_exp}
\end{equation}
We recall that according to the MWC theory, the density at the grafting
surface is defined solely by the total grafting density and the number-averaged
chain length, irrespective of the polydispersity, $\phi(0)=A$. Recognizing
that $U/A=1-\phi/\phi(0)$, $H_{\mathrm{mono}}=\frac{2}{\pi}\sqrt{2A/3}N_{n}$,
one can express the density profile as: 
\begin{equation}
\phi_{\mathrm{exp}}(z)=\phi(0)\cdot\Psi\left(\frac{z}{3H_{\mathrm{mono}}}\right),\label{eq:exp_distr_profile-1}
\end{equation}
where $y=\Psi(x)$ is the inverse of the function $x=\sqrt{1-y}-\sqrt{y}\arcsin(\sqrt{1-y})$.
The density profile terminates at the distance $z=3H_{\mathrm{mono}}$.
The end monomer position as a function of the chain length is found
by substituting $U(N)$ from \eqn (\ref{eq:UN_exp}) into \eqn (\ref{eq:zU_exp})
and identifying $z(U(N))$ with $Z_{e}(N)$, which serves approximately
as the average position of the free end monomer.

\section{Polydisperse Brush in Hydrodynamic Shear Flow}

\label{sec:appendix_flow}

The solution of the Brinkman equation (\eqn (\ref{eq:brinkman})) with $\lambda=\phi^{-1}$
and the density profile given by \eqn (\ref{eq:miln_lin_profile}) for
monodisperse brush was obtained by Milner \cite{Milner:hydrodynamic}
and has the form
\begin{equation}
v(x)=\frac{v(0)}{\Gamma(1/4)}\left(2x\sqrt{\frac{2\phi_{0}}{H_{\mathrm{mono}}}}\right)^{1/2}K_{1/4}\left(x^{2}\frac{\phi_{0}}{H_{\mathrm{mono}}}\right)\label{eq:solution_miln_mono-1}
\end{equation}
where $K$ is the modified Bessel function, and $\Gamma$ is the
Euler gamma function. The numerical prefactor in the expression for
the penetration depth is obtained from the initial slope of the function.

For the strongly polydisperse brush with $\pdi=2$, i.e., 
the density profile given by \eqn (\ref{eq:tail_strong_poly}), we introduce
the dimensionless variable 
$y=x\phi_{0}^{1/3}\left(\frac{2}{3\pi H_{\mathrm{mono}}}\right)^{2/3}$. 
The Brinkman equation (\ref{eq:brinkman}) then reduces to: 
\begin{equation}
\frac{d^{2}v}{dy^{2}}=y^{4}v\label{eq:quadratic_profile},
\end{equation}
which has an exact solution of the form 
\begin{equation}
v(y)=v(0)\,\frac{\Gamma(5/6)}{\pi6^{1/6}}\, K_{1/6}\left(\frac{y^{3}}{3}\right)\label{eq:quad_solution}
\end{equation}


\begin{thebibliography}{99}
\section*{References}
\bibitem{Alexander:1977} Alexander, S. J. Adsorption of chain molecules with a polar head a scaling description. \textit{J. Phys. France} {\bf 1977}, 38, 983-987.
\bibitem{Biofouling}Bixler, G. D.; Bhushan B. Biofouling: lessons from nature. \textit{Philos. Trans. R. Soc. A} {\bf 2012}, 370, 2381-2417.
\bibitem{Klein:1991}Klein, J.; Perahia, D.; Warburg, S. Forces between polymer-bearing surfaces undergoing shear. \textit{Nature} {\bf 1991}, 352, 143-145.
\bibitem{Singh:2015}Singh, M. K.; Ilg, P.; Espinosa-Marzal, R. M.; Kr\"oger, M.; Spencer, N. D. Polymer brushes under shear: molecular dynamics simulations compared to experiments. \textit{Langmuir} {\bf 2015}, 31, 4798-4805.
\bibitem{Zdyrko:2009}Zdyrko, B.; Klep, V.; Li, X.; Kang, Q.; Minko, S.; Wen, X.; Luzinov, I. Polymer brushes as active nanolayers for tunable bacteria adhesion. \textit{Materials Science and Engineering: C} {\bf 2009}, 29, 680-684.
\bibitem{Ayres:2010}Ayres, N. Polymer brushes: applications in biomaterials and nanotechnology. \textit{Polym. Chem.} {\bf 2010}, 1, 769-777.
\bibitem{Stimuli_responsive}Marek, U. U. \textit{Handbook of Stimuli-Responsive Materials}; Viley-VCH Verlag GmbH \& Co. KGaA: Weinheim, Germany, 2001.
\bibitem{Cohen_Stuart}Cohen Stuart, M. A.; Huck, W. T. S.; Genzer, J.; M\"uller, M.; Ober, C.; Stamm, M.; Sukhorukov, G. B.; Szleifer, I.; Tsukruk, V. V.; Urban, M.; Winnik, F.; Zauscher, S.; Luzinov, I.; Minko, S. Emerging applications of stimuli-responsive polymer materials. \textit{Nat. Mater.} {\bf 2010}, 9, 101-113.
\bibitem{MWC}Milner, S.; Witten, T.; Cates, M. Theory of grafted polymer brush. \textit{Macromolecules} {\bf 1988}, 21, 2610-2619.
\bibitem{Zhulina}Zhulina, E. B.; Borisov, O. V.; Priamitsyn, V. A. Theory of steric stabilization of colloid dispersions by grafted polymers. \textit{Coll. Interface Sci.} {\bf 1990}, 137, 495-511.
\bibitem{Skvortsov:1988}Skvortsov, A. M.; Pavlushkov, I. V.; Gorbunov, A. A.; Zhulina, E. B.; Borisov, O. V.; Priamitsyn, V. A. Structure of densely grafted polymeric monolayers. \textit{Vysokomol. Soedin. A} {\bf 1988}, 30, 1598.
\bibitem{Advincula:2011}Advincular, R. C.; Brittain, W. J.; Caster, K. C.; Ruhe, J. \textit{Polymer Brush}; Viley-VCH, Weinheim, Germany, 2011.
\bibitem{Binder_Kreer:2011}Binder, K.; Kreer, T.; Milchev, A. Polymer brushes under flow and in other out-of-equilibrium conditions. \textit{Soft Matter} {\bf 2011}, 7, 7159-7172.
\bibitem{Halperin:1992}Halperin, A.; Tirrell, M.; Lodge, T. P. Tethered chains in polymer microstructures. \textit{Adv. Polym. Sci.} {\bf 1992}, 100, 31-71.
\bibitem{Romeis:2012}Romeis, D.; Merlitz, H.; Sommer, J.-U. A new numerical approach to dense polymer brushes and surface instabilities. \textit{J. Chem. Phys.} {\bf 2012}, 136, 044903.
\bibitem{Netz:2006}Naji, A.; Seidel. C.; Netz, R. R. Theoretical approaches to neutral and charged polymer brushes. \textit{Adv. Polym. Sci.} {\bf 2006}, 198, 149-183.
\bibitem{Netz:1998}Netz, R. R.; Schick, M. Polymer brushes: from self-consistent field theory to classical theory. \textit{Macromolecules} {\bf 1998}, 31, 5105-5122.
\bibitem{Minko2016}Laradji, A. M.; McNitt, C. D.; Yadavalli, N. S.; Popik, V. V.; Minko, S. Roubust, solvent-free, catalyst-free click chemistry for the generation of highly stable densely grafted poly(ethylene glycol) polymer brushes by the grafting to method and theory properties. \textit{Macromolecules} {\bf 2016}, 49, 7625.




\bibitem{Turgman}Turgman, C. S.; Srogl, J.; Kiserow, D.; Genzer, J. On-demand degrafting and the study of molecular weight and grafting density of Poly(methyl methacrylate) brushes on flat silica substrates. \textit{Langmuir} {\bf 2015}, 31, 2372-2381.
\bibitem{Pandav}Pandav, G.; Ganesan, V.; Fluctuation effects on the order-disorder transition in polydisperse copolymer melts. \textit{J. Chem. Phys.} {\bf 2013}, 139, 214905.
\bibitem{Balko:2013}Balko, S. M.; Kreer, T.; Costanzo, P. J.; Patten, T. E.; Johner, A.; Kuhl, T. L.; Marques, C. M. Polymer brushes under high load. \textit{PloS ONE} {\bf 2013}, 8(3): e58392.
\bibitem{Russel:2007}Murata, H.; Koepsel, R. R.; Matyjaszewski, K.; Russell, A. J. Permanent, non-leaching antibacterial surfaces-2: How high density cationic surfaces kill bacterial cells. \textit{Biomaterials} {\bf 2007}, 28, 4870-4879.
\bibitem{Krishnamoorthy:2014}Krishnamoorthy, M.; Hakobyan, S.; Ramstedt, M.; Gautrot, J. E. Surface-initiated polymer brushes in the biomedical field: applications in membrane science, biosensing, cell culture, regenerative medicine and antibacterial coatings. \textit{Chem. Rev.} {\bf 2014}, 114, 10976-11026.

\bibitem{Milner_Polydisp}Milner, S.; Witten, T.; Cates, M. Effects of polydispersity in the end-grafted polymer brush. \textit{Macromolecules} {\bf 1989}, 22, 853-861.
\bibitem{Leermakers}de Vos, W. M.; Leermakers, F. A. M. Modeling the structure of a polydisperse polymer brush. \textit{Polymer} {\bf 2009}, 50, 305-316.
\bibitem{deVos_nanopartcle}de Vos, W. M.; Leermakers, W. M.; de Keizer, A.; Cohen Stuart, M. A. Interaction of particles with a polydisperse brush: A self-consistent-field analysis. \textit{Macromolecules} {\bf 2009}, 42, 5881-5891.


\bibitem{EdwardsType1}Helfand, E. Theory of inhomogeneous polymers: Fundamentals of the Gaussian random‐walk model. \textit{J. Chem. Phys.} {\bf 1975}, 62, 999-1005
\bibitem{EdwardsType2}Laradji M.; Guo H.; Zuckermann M. J. Off-lattice Monte Carlo simulation of polymer brushes in good solvents. \textit{Phys. Rev. E} {\bf 1994}, 49, 3199-3206.
\bibitem{EdwardsType3}Besold, G.; Guo, H.; Zuckermann, M. J. Off-lattice Monte Carlo simulation of the discrete Edwards model. \textit{J. Polym. Sci. Part B: Polym. Phys.} {\bf 2000}, 38, 1053-1068.
\bibitem{Hybrid_PF}Qi, S.; Behringer, H.; Schmid, F. Using field theory to construct hybrid particle-continuum simulation schemes with adaptive resolution for soft matter systems. \textit{New J. Phys.} {\bf 2013}, 15, 125009.
\bibitem{Klushin:2015}Qi, S.; Klushin, L. I.; Skvortsov, A. M.; Polotsky, A. A.; Schmid, F. Stimuli-responsive brushes with active minority components: Monte Carlo study and analytical theory. \textit{Macromolecules} {\bf 2015}, 48, 3775-3787.
\bibitem{chi_v}Milner, S. T.; Lacasse, M.-D.; Graessley, W. M. Why $\chi$ is seldom zero for polymer-solvent mixtures. \textit{Macromolecules} {\bf 2009}, 42, 876-886. 
\bibitem{Particle_Mesh}Detcheverry, F. A.; Kang, H.; Daoulas, K.; M\"uller, M.; Nealey, P. F.; de Pablo, J. J. Monte Carlo simulations of a coarse Grain model for block copolymers and nanocomposites. \textit{Macromolecules} {\bf 2008}, 41, 4989-5001
\bibitem{CIC} Birdsall, C.K.; Fuss, D. Clouds-in-clouds, clouds-in-cells physics for many-body plasma simulation. \textit{J. Comput. Phys.} {\bf 1997}, 135, 141-148.
\bibitem{Schulz}Schulz, G. V. Z. \textit{Z. Phys. Chem.} (Munich) {\bf 1939}, B43, 25-46.
\bibitem{Zimm}Zimm, B. Apparatus and methods for measurement and interpretation of the angular variation of light scattering; preliminary results on polystyrene solutions. \textit{J. Chem. Phys.} {\bf 1948}, 16, 1099-1116.
\bibitem{SCF_book}Fredrickson, G. H. {\bf 2006} \textit{The Equilibrium Theory of Inhomogeneous Polymers} (Oxford: Oxford University Press)

\bibitem{Binder_Milchev_2012}Binder, K.; Milchev, A.; Polymer brushes on flat and curved surfaces: How computer simulations can help to test theories and to interpret experiments. \textit{J. Polym. Sci. Part B: Polym Phys} {\bf 2012}, 50, 1515-1555. 
\bibitem{Klushin:1992}Klushin, L. I.; Skvortsov, A. M. Chain behavior in a polydisperse brush: Depression of critical fluctuations. \textit{Macromolecules} {\bf 1992}, 25, 3443-3448.
\bibitem{Mansfield}Mansfield, M. L. The coil–stretch transition of polymers in external fields. \textit{J. Chem. Phys.} {\bf 1988}, 88, 6570-6580.
\bibitem{Skvortsov:1997}Skvortsov, A. M.; Gorbunov, A. A.; Klushin, L. I. Long and short chains in a polymeric brush: A conformational transition. \textit{Macromolecules} {\bf 1997}, 30, 1818-1827.


\bibitem{Klushin:1991}Klushin, L. I.; Skvortsov, A. M. Critical dynamics of a polymer chain in a grafted monolayer. \textit{Macromolecules} {\bf 1991}, 24, 1549-1553.
\bibitem{Polymer_solution}des Cloizeaux, J.; Jannink, G; {\bf 1990} \textit{Polymers in solution:
their modelling and structure} (Claradon, Oxford)
\bibitem{Lai_Zhulina_1992}Lai, P.-Y.; Zhulina, E. B. Monte Carlo test of the self-consistent field theory of a polymer brush. \textit{J. Phys. II France 2} {\bf 1992} 2. 547-560.

\bibitem{Lai:1991}Lai, P. Y.; Binder, K. Structure and dynamics of grafted polymer layers: A Monte Carlo simulation. \textit{J. Chem. Phys.} {\bf 1991}, 95, 9288-9299.
\bibitem{Lai:1994}Binder, K.; Lai, P. Y.; Wittmer, J. Monte Carlo simulations of chain dynamics in polymer brushes. \textit{Faraday Discussion} {\bf 1994}, 98, 97-109.
\bibitem{Marko:1993}Marko, J. F.; Chakrabarti, A. Static and dynamic collective correlations of polymer brushes.\textit{Phys. Rev. E} {\bf 1993}, 48, 2739-2743.
\bibitem{Virnau:2012}Reith, D.; Milchev, A.; Virnau, P.; Binder, K. Computer simulation studies of chain dynamics in polymer brushes. \textit{Macromolecules} {\bf 2012}, 45, 4381-4393.

\bibitem{Klushin:2014} Klushin, L.I.; Skvortsov, A. M.; Polotsky, A. A.; Qi, S.; Schmid, F. Sharp and fast: Sensors and switches based on polymer brushes with adsorption-active minority chains. \textit{Phys. Rev. Lett.} {\bf 2014}, 113, 068303.
\bibitem{Merlitz:2008}Merlitz, H.; He, G.-L.; Wu, C.-V.; Sommer, J.-U. Surface instabilities of monodisperse and densely grafted polymer brushes. \textit{Macromolecules} {\bf 2008}, 41, 5070-5072.
\bibitem{Romeis:2013} Romeis, D.; Sommer, J.-U. Conformational switching of modified guest chains in polymer brushes. \textit{J. Chem. Phys.} {\bf 2013}, 139, 044910.
\bibitem{Romeis:2015} Romeis, D.; Sommer, J.-U. Binary and bidisperse polymer brushes: coexisting surface states. \textit{ACS Appl. Mater. Interfaces} {\bf 2015}, 7, 12496-12504.

\bibitem{Doyle:1998}Doyle, P. S.; Shaqfeh, E. S. G.; Gast, A. P. Rheology of polymer brushes:  A brownian dynamics study. \textit{Macromolecules} {\bf 1998}, 31, 5474-5486.



\bibitem{Milner:hydrodynamic}Milner, S. Hydrodynamic penetration into parabolic brushes. \textit{Macromolecules} {\bf 1991}, 24, 3704-3705.
\bibitem{Klein:1996}Klein, J. Shear, friction, and lubrication forces between polymer-bearing surfaces. \textit{Annu. Rev. Mater. Sci.} {\bf 1996}, 26, 581-612.
\bibitem{Brinkman:1947}Brinkman, H. C. A calculation of the viscous force exerted by a flowing fluid on a dense swarm of particles. \textit{Appl. Sci. Res.} {\bf 1947}, A1, 27-34.
\bibitem{Lai:1993}Lai, P.-Y.; Binder, K. Grafted polymer layers under shear: A Monte Carlo simulation. \textit{J. Chem. Phys.} {\bf 1993}, 98, 2366-2375.
\bibitem{Whitmore:2014}Suo, T.; Whitmore, M. D. Doubly self-consistent field theory of grafted polymers under simple shear in steady state. \textit{J. Chem. Phys.} {\bf 2014}, 140, 114901.
\bibitem{Wijmans:2002}Wijmans, C. M.; Smit, B. Simulating tethered polymer layers in shear flow with the dissipative particle dynamics technique. \textit{Macromolecules} {\bf 2002}, 35, 7138-7148.
\bibitem{Deng:2012}Deng, M.; Li, X.; Liang, H.; Caswell, B.; Karniadakis, G. Em. Simulation and modelling of slip flow over surfaces grafted with polymer brushes and glycocalyx fibres. \textit{J. Fluid Mech.} {\bf 2012}, 711, 192-211.
\bibitem{entanglement_brush}Lang, M.; Werner, M.; Dockhorn, R.; Kreer, T.; Arm retraction dynamics in dense polymer brushes. \textit{Macromolecules} {\bf 2016}, 49, 5190-5201.


\bibitem{SCF_R1} Schmid F. Self-consistent-field theories for complex fluids. \textit{J. Phys.: Condens. Matter} {\bf 1998}, 10, 8105-8136.
\bibitem{numerical_recipes}Press, W. H.; Teukolsky, S. A.; Vetterling, W. T.; Flannery, B. P. \textit{Numerical Recipes in C: The Art of Scientific Computing}; 2nd Edition, Cambridge University Press: 1992

\bibitem{Feynman}Feynman, R. P.; Hibbs, A. R.; Stayer, D. F. {\bf 2005} \textit{Quantum Mechanics and Path Integrals} (New York: MCGraw Hill)




\end{thebibliography}
\end{document}